\documentclass[12pt,tightenlines,eqsecnum,floats,aps,amsmath,amssymb,nofootinbib,showpacs]{revtex4}

\usepackage{amsmath,amsthm,amssymb,amsfonts}

\usepackage{dcolumn}
\usepackage{bm}
\usepackage[latin1]{inputenc}
\usepackage[spanish,english]{babel}
\usepackage{amsfonts}
\usepackage{amssymb}
\usepackage{graphicx}
\usepackage{verbatim}
\usepackage{color}

\newcommand{\be}{\begin{equation}}
\newcommand{\ee}{\end{equation}}
\newcommand{\bea}{\begin{eqnarray}}
\newcommand{\eea}{\end{eqnarray}}

\begin{document}

\title{Non-gaussianities and the Stimulated creation of quanta in the inflationary universe}

\author{Ivan Agullo}
\affiliation{ {\footnotesize Physics Department, University of
Wisconsin-Milwaukee, P.O.Box 413, Milwaukee, WI 53201 USA}}

\author{Leonard Parker}
\affiliation{ {\footnotesize Physics Department, University of
Wisconsin-Milwaukee, P.O.Box 413, Milwaukee, WI 53201 USA}}

\date{August 27, 2010}

\begin{abstract}

Cosmological inflation generates a spectrum of density perturbations that can seed the cosmic structures we observe today. These perturbations are usually computed as the result of the gravitationally-induced spontaneous creation of perturbations from an initial vacuum state. In this paper, we compute the perturbations arising from gravitationally-induced stimulated creation when perturbations are already present in the initial state. The effect of these initial perturbations is not diluted by inflation and survives to its end, and beyond.  We consider a generic statistical density operator $\rho$ describing an initial mixed state that includes probabilities for nonzero numbers of scalar perturbations to be present at early times during inflation. 
We  analyze the primordial bispectrum for general configurations of the three different momentum vectors in its arguments. We find that the initial presence of quanta can significantly enhance non-gaussianities in the so-called squeezed limit.  Our results show that an observation of non-gaussianities in the squeezed limit can occur for single-field inflation when the state in the very early inflationary universe is not the vacuum, but instead contains early-time perturbations. Valuable information about the initial state can then be obtained from observations of those non-gaussianities.

\end{abstract}


\maketitle

\section{Introduction} \label{introduction}

Quantum field theory in curved spacetimes can be considered at the present time as a well established framework (a recent introduction can be found in \cite{parker-toms} and references cited there).  In this framework, quantum theory and the effects of general relativity are brought together in the regime where we are confident about the validity of both theories. One studies, then, the behavior of quantum fields propagating in a spacetime described by a classical metric as in general relativity. In that way, the fundamental difficulty of formulating quantum field theory in the absence of a spacetime background and treating the quantum aspects of nonlinear gravity is avoided. However, in spite of its semiclassical  character,  very profound physical results have been obtained in quantum field theory (QFT) in curved spacetimes. One of the most remarkable outcomes of this framework is the phenomenon of spontaneous creation of quanta by curved spacetimes, as first pointed out and analyzed in \cite{parkerthesis, parker68-69-71} in the cosmological context of an expanding universe. Fundamental implications of this phenomenon have been obtained in physical situations where spacetime curvature plays an important role, namely in the vicinity of black holes and in the early universe. In the former case, Hawking predicted the spontaneous emission of thermal radiation by black holes that are produced by gravitational collapse \cite{hawk1}. In the cosmological scenario, when the idea of inflation was investigated \cite{inflation},  it was predicted  \cite{inflation2} that the inflationary expansion of the universe is able to spontaneously create a density of perturbations and gravitational waves with a nearly scale-invariant spectrum. Black hole thermal radiance has not been detected yet, implying a low density of radiating primordial black holes in the observable universe.  On the other hand, the high degree of precision in measurements of the temperature fluctuations of the CMB and the large scale structure (LSS) achieved during the last decade, together with the new generation of observational missions, such as the PLANCK satellite, open the possibility of scrutinizing in great detail the possible effects of the physics of the early universe. In particular, in the inflationary scenario the temperature fluctuations of the CMB and the LSS are directly related to the spectrum of density perturbations created during inflation. Therefore, the observation of the CMB and LSS open a fascinating window to measure direct consequences of the predictions of QFT in curved spacetimes, converting cosmology into a suitable scenario where this fundamental physics can be tested. 

The present observation of the CMB temperature inhomogeneities \cite{wmap7} indicates that they are distributed following a nearly scale-invariant spectrum. They thus are compatible with the predictions of inflation. The new generation of observational missions will significantly improve the precision of the measurements of CMB temperatures and polarizations, and of the LSS,  in such a way that they may constitute a test of the inflationary paradigm and could provide enough information to favor some specific models for inflation and rule out others. In particular, the study of the non-gaussianities of the temperature fluctuations of the CMB and LSS will offer a sharp tool to obtain specific information about the dynamics of the field or fields driving inflation (see \cite{whitepaper} and references therein). 

The predictions of inflation for the spectrum of density perturbations produced at the end of inflation are generally computed by studying the spontaneous generation of quanta from a suitable vacuum state (see for example \cite{Weinberg2008,dodelson,liddlelyth,baumann}). Such a state for describing the inflaton quantum fluctuations, in the context of slow-roll inflation, is chosen as the natural extension of the Bunch-Davies vacuum \cite{bunch-davies}. This vacuum state is compatible with the symmetries of the spacetime in the limit where the spacetime geometry is exactly de Sitter. The election of the vacuum as the state describing the inflaton quantum fluctuations at the beginning of inflation can seem unnatural due to our ignorance of the physics in the early universe before inflation. However, it is consistent with the argument that the exponential expansion of the universe during inflation will dilute any possible quanta present in the initial state and will drive any arbitrary state to the Bunch-Davies vacuum. That justifies the vacuum state as the most natural choice. But generically, the phenomenon of spontaneous emission of quanta is accompanied by the corresponding stimulated or induced counterpart. As first pointed out in \cite{parkerthesis,parker68-69-71}, the gravitationally-induced spontaneous creation of quanta in a general  expanding universe is accompanied by the corresponding stimulated process if there are quanta already present in the initial state, which most generally could be a mixed state described by a quantum mechanical density operator, $\rho$. The objective of this paper is to study the effects of the stimulated creation of quanta when a general mixed state, $\rho$, describes the inflaton quantum fluctuations at the beginning of inflation.

The stimulated particle creation process (as may be seen from \cite{parkerthesis, parker68-69-71}) is such that the effects of the modes that are occupied survive to the end of inflation and beyond. The energies and momenta of the initial occupied modes are considerably red-shifted by the present time. Nevertheless, the common argument that an initial density of particles present before inflation started would become negligible after many $e$-foldings of inflation is not correct. On the contrary, the stimulated creation process during inflation is enhanced by the same quantum amplification factor as is responsible for the spontaneous particle creation from the vacuum that gives rise to the non-negligible amplitude of perturbations present after inflation. For the same reason that the effects of the quantized field fluctuations are observable in the spectrum of perturbations of the CMB and in the LSS, the effects of quanta present at the start of inflation are not dispersed away by inflation and may possibly give rise to observable consequences today.

In this paper, we start with a general mixed quantum state $\rho$ describing a statistical distribution of initial inflaton perturbations parametrizing our ignorance about  the physics before inflation. Such initial perturbations would appear to be plausible in models in which our universe originated through inflation from a much larger (possibly inhomogeneous and non isotropic) universe. Previous works have considered  the possibility of having a non-vacuum initial state by considering a thermal state at a given temperature \cite{gasperiniveneziano,bhattacharya,magueijo} or an exactly pure state given by a Bogoliubov transformation of the Bunch-Davies vacuum \cite{boyanovsky,holman,meerburg} similar to the state first considered in \cite{parkerfulling73}. Here we consider a more generic mixed state, described by a statistical density operator formed from individual pure states having definite initial numbers of perturbation quanta (i.e., particles). The initial density operator need not necessarily be isotropic, within the observational constraints. The condition that the contribution to the energy density does not significantly influence the inflating background geometry largely restricts the number of initial quanta. We compute the expression for the two-point function in momentum space (i.e., the power spectrum) for scalar curvature perturbations in our general mixed initial state and find constraints on the state that follow from the observed values of the amplitude and spectral index of the scalar power spectrum. We then compute the complete expression for the three-point function in momentum space (the bispectrum) using the canonical Lagrangian for curvature perturbations. We find that the three-point function is much more sensitive than the two-point function to the initial state, in agreement with previous analyses \cite{holman,meerburg,bhattacharya}. 
In particular, we obtain a significant enhancement of the non-gaussianties for certain configurations of the three momenta involved in the three point function when the number of initial quanta is equal to or greater than $1$.

However, in contrast with previous analyses, we find a significant enhancement of the non-gaussianities for the configuration in which two of the momenta that appear in the bispectrum are much bigger than the third one, for instance $k_1\approx k_2\gg k_3$ (the so-called squeezed configuration). In this configuration, the value of the $f_{NL}$ parameter that characterizes the level of non-gaussianities \cite{komatsuspergel,komatsuthesis} can be enhanced by a factor $k_1/k_3$ with respect to its value when the pure vacuum state is used. This enhancement factor can be as large as a few hundred for the scales accessible in observations.  Our result means that if non-gaussianities are found to be observable in the squeezed configuration, then single-field inflation is not necessarily ruled out, as would follow from \cite{creminellizaldarriaga}.  
We show that such an observation could be interpreted as a consequence of an initial state for inflation that contains initial perturbations that have effects observable today because of the stimulated production of quanta during inflation.

The rest of the paper is organized as follows. Section II is devoted to the introduction of the mixed initial state that we use in this paper and to the restriction that gravitational backreaction imposes on it. In section III, we give a pedagogical introduction to the phenomenon of stimulated creation of perturbations in an expanding universe. In section IV, we work out the expression for the power spectrum of scalar perturbations when the initial mixed state contains non-zero numbers of quanta. In section V, we present the computation of the  three-point function and compute the level of non-gaussianities produced in different configuration. We finish in section VI with some conclusions and final remarks.

\section{Choice of state at an initial time \label{initialstate}}

In the context of inflation, the density and temperature fluctuations that we observe at a given scale in the present universe originated at the time when the modes corresponding  to that scale left the Hubble radius during inflation. The modes that are just coming into our observable universe at the present time exited the Hubble radius around 60 $e$-foldings before the end of inflation. Any shorter wavelengths observable today would have exited the Hubble radius at later times during inflation. Therefore, around 60 $e$-foldings is the earliest time that we can explore through observation of the non-unifomities of the present universe. We take that time (or a few $e$-foldings earlier) as the time $t_0$ for setting up initial conditions for the quantum state of the inflaton perturbations that we want to study. At the time $t_0$, the energy density of the universe was some orders of magnitude below the Planck scale (see for example \cite{Weinberg2008} for details) and the modes that we observe today in the inhomogeneities of the CMB and LSS were sub-planckian at that time. Therefore, we do not need to consider the excitation of trans-planckian modes for studying the generation of the non-uniformities that we observe today, and we can use QFT in curved spacetime as a plausible effective field theory. 

As discussed earlier, the state of the inflaton quantum fluctuations at this initial time $t_0$ is necessary to compute the density of perturbations produced during inflation that eventually give rise to the temperature fluctuations of the CMB and to the LSS.  In this situation, it is natural to proceed by writing the most general initial state for inflaton fluctuations using a reasonable number of variables parameterizing our ignorance about the initial state. Then, we can use the observational data to constrain these variables. 

We describe the state of the fluctuation field at the time $t_0$ by means of a statistical density operator $\rho$. The possibility of states having nonzero numbers of fluctuation quanta in the modes at $t_0$ will be included, along with the possibility of the vacuum state, in the operator $\rho$.  Such a mixed state at $t_0$ is natural within the context of inflation. In inflation, our observable universe is a portion of the region of space that results from the enormous expansion of a small patch of the initial universe. The state of such a patch would naturally be a mixed state because many features of the global state of the whole universe would not be accessible to our observable universe. As described in \cite{vonneumann}, the state of our subsystem (i.e., our observable universe) arises from tracing out the full density operator of the large universe over ``unobservables'' from the point of view of the patch giving rise to our observable universe. That state would be a mixed state described by a density operator, even if the global state of the larger universe were a pure state.  In such a scenario, the initial density operator of our universe need not be that of an equilibrium canonical or grand canonical ensemble.

Motivated by such considerations, we take a general density operator $\rho$ formed from individual pure states having definite initial numbers\footnote{ A mixed state formed from coherent states having indefinite numbers of initial particles would also be of interest, but we limit ourselves here to the former type of mixed state.}
of perturbation quanta (i.e., particles)
\begin{equation}
\rho =  \sum_{n_1=0}^{\infty}\sum_{n_2=0}^{\infty}\cdots\, p( n_1(j_1), n_2(j_2),\ldots )  \, 
|n_1(j_1), n_2(j_2),\ldots\rangle \,\langle n_1(j_1), n_2(j_2),\ldots | \ ,
\label{rho1}
\end{equation}
where the complete orthonormal set of basis states is given by the states
\begin{equation} 
|n_1(j_1), n_2(j_2),\cdots\rangle \equiv
(n_1! n_2! \cdots)^{-1/2}
(A_{j_1}{}^\dagger)^{n_1}(A_{j_2}{}^\dagger)^{n_2}\cdots |0\rangle,
\label{rho2}
\end{equation}
containing $n_1$ quanta in mode $j_1$, $n_2$ quanta in mode $j_2$, \ldots, and the probabilities $p$ sum to $1$
\begin{equation} 
\sum_{n_1=0}^{\infty}\sum_{n_2=0}^{\infty}\cdots\, p( n_1(j_1), n_2(j_2),\ldots ) = 1.
\label{rho3}
\end{equation}
In arriving at the above expression for $\rho$, the quantized scalar perturbation field, call it $\varphi$, has been expanded as
\be
\varphi ({\vec x},t)=\sum_j \left(A_j \varphi_j({\vec x},t) + A_j{}^\dagger \varphi_j{}^*({\vec x},t)\right),
\label{rho4}
\ee
in terms of a complete orthonormal set of wave packet solutions, $\varphi_j(x)$, of the perturbation field equation. The index $j$ is discrete, and the annihilation and creation operators for quanta of the perturbation field satisfy the commutation relations
\be 
[A_i,A_j{}^\dagger] = \delta_{i,j}, \hspace{2mm} [A_i,A_j] = 0 \ ,
\label{rho5}
\ee
where $ \delta_{i,j}$ is the Kronecker delta.

The expectation value of any Hermitian observable operator $B$ formed from $\varphi$ and its conjugate momentum is given by
\be
\langle B \rangle = {\rm Tr}[\rho B] = \sum_{n_1=0}^{\infty}\cdots\, p( n_1(j_1),\ldots)  \, \langle n_1(j_1),\ldots | \, B\, | n_1(j_1), \ldots \rangle\ .
\label{rho6}
\ee
It is clear from (\ref{rho6}) that, unless the operator $B$ contains terms that have an equal number of creation and annihilation operators, $ A_j{}^\dagger$ and  $A_j$, for each mode, the expectation value of $B$ will vanish.

In the Heisenberg picture, the vector states and hence $\rho$, do not evolve, but operators like $B$ carry the time dependence. The reverse is true in the Schr{\"o}dinger picture, so that $\rho$ alone carries the time-dependence. In the interaction picture, the vector states are evolved by the interaction part of the Hamiltonian and the measured observables like $B$, are evolved by the unperturbed part of the Hamiltonian.

The information encoded in $\rho$ that will be relevant for the computation of the power spectrum and the bispectrum of scalar perturbations in the following sections is the average numbers of initial quanta present in a given mode, that is, the quantities like ${\rm Tr}[\rho N_{\vec k}]$, where $N_{\vec k}$ is the operator for the number of quanta in mode with coordinate momentum ${\vec k}$. Here, we have written ${\vec k}$ because the discrete set of modes we use in next sections are plane waves obeying periodic boundary conditions. 

However, before considering the direct observational constraints on ${\rm Tr}[\rho N_{\vec{k}} ]$, it is important to note (see \cite{boyanovsky,holman}) that strong conditions can be imposed over these average numbers of particles, or quanta of the perturbation field, on theoretical grounds. On the one hand, an acceptable initial state must insure that physical quantities such as the energy-momentum tensor do not diverge when using standard methods of renormalization in curved spacetimes (see, for instance, \cite{parker-toms}). That condition restricts the behavior of the function ${\rm Tr}[\rho N_{\vec{k}} ]$ for large values of $k\equiv |\vec{k}|$ to fall-off as $\sim 1/k^{4+\delta}$, with $\delta>0$. 

A more restrictive condition for the value of ${\rm Tr}[\rho N_{\vec{k}} ]$ comes from the requirement that backreaction effects from the perturbations present at $t_0$ should not modify the inflationary dynamics described by the non-perturbative part of the inflaton field. That implies that the contribution to the energy density of the inflaton perturbations at $t_0$ has to be negligible compared to the energy density of the unperturbed part of the inflaton field, which is given by the inflaton potential $V(\phi)\sim M_P^2 H^2$. One can obtain an estimate of the restrictions that this condition imposes on the value of ${\rm Tr}[\rho N_{\vec{k}} ]$ by considering the crude approximation ${\rm Tr}[\rho N_{\vec{k}} ]\approx N$ for all $k/a(\tau_0)<\mu$, where $N$ is a constant, and ${\rm Tr}[\rho N_{\vec{k}} ]=0$ for all $k/a(t_0)>\mu$, where $\mu$ can be considered as the energy scale up to which the effective theory that we are considering is valid. The initial state energy density is then
\be \label{energydensity1} \frac{1}{a^4(t)}\int_0^{\mu a(t_0)} d^3k k N \sim \frac{a^4(t_0)}{a^4(t)} \mu^4 N \lesssim M_P^2 H^2 \ . \ee
The strongest restriction on $N$ coming from this inequality is obtained when $t=t_0$, when the previous condition requires that $N\lesssim M_P^2 H^2/\mu^4$. For the reasonable value $\mu\sim\sqrt{M_P H}\sim V(\phi)^{1/4}$, it follows that $N$ can be of order one. If we take $\mu\sim H$, larger values of $N$ are allowed. A more detailed analysis, including effects of renormalization, can be done \cite{boyanovsky} (see also \cite{anderson}), and similar results, up to factors of order one, are obtained. Thus, renormalization and backreaction impose conditions on the average number of particles at $t_0$, but do not necessarily restrict us to small values of ${\rm Tr}[\rho N_{\vec{k}} ]$ for all $\vec{k}$. In sections \ref{powspec} and \ref{bispec} we derive expressions for the power spectrum and bispectrum, respectively, of scalar perturbations present during inflation for an initial state described by a density operator $\rho$ of the type described in this section. The resulting expressions for the spectrum and bispectrum can be directly compared with observational data to impose further restrictions on ${\rm Tr}[\rho N_{\vec{k}} ]$. As we will see, while the power spectrum imposes minor constraints, the bispectrum is a sharp tool to restrict the initial state for scalar perturbations.

\section{Stimulated creation of perturbations in an expanding universe \label{stimulated}}

The effect of the stimulated creation of particles or perturbations of a scalar field was first derived in an expanding universe in \cite{parkerthesis,parker68-69-71}, where the effect was given for the particle number of a quantized scalar field, or equivalently, for the number of quanta of a scalar perturbation field. (As noted earlier, we can refer to quanta of the perturbation field as particles of that field, since they are governed by similar linear wave equations.)
Here we focus on the dispersion of the scalar field perturbations .  For readers who may not be familiar with this early work, we find it useful in this section to give a pedagogical introduction to the phenomenon of stimulated creation of perturbations in an expanding universe.

Consider a spatially flat Friedmann-Robertson-Walker spacetime given by the line element

\be \label{FRW} ds^2=dt^2-a^2(t) (dx^2+dy^2 +dz^2) \ ,\ee
where $a(t)$ is the scale factor. In this geometry let us consider, for simplicity and for physical interest (next sections), a free, massless, minimally coupled scalar field $\varphi(\vec{x},t)$  (the generalization to the massive case and the inclusion of arbitrary coupling to the Ricci scalar curvature is straightforward) satisfying the Klein-Gordon equation
\be \label{KG} \Box \varphi(\vec{x},t)=0 \ . \ee

Let us proceed to quantize the field $\varphi(\vec{x},t)$.  In general, the statistical density operator $\rho$ is a positive definite Hermitian operator in a Hilbert space.  Then $\rho$ must have a discrete set of eigenvalues \cite{messiah}.  Indeed, the set of probabilities $p( n_1(j_1),\ldots)$ (i.e., the eigenvalues of $\rho$) defined in the previous section are discrete.  We find it convenient to employ spatially periodic boundary conditions in the quantization, such that $\varphi(\vec{x}+\vec{n} L,t)=\varphi(\vec{x},t)$, where $\vec{n}\in \mathbb{Z}^3$. The use of periodic boundary conditions introduces discreteness in the values of the momenta $\vec{k}=2\pi L^{-1} \vec{n}$. Working with discrete momenta also has advantages at the technical level, and in the fact that the interpretation of quantities like the number operator is straightforward compared to the continuum case. One can also regard the periodic boundary conditions as a convenient mathematical device, such that after quantities of physical interest have been calculated, one can take the limit $L\to\infty$.\footnote{The continuum limit can be obtained by using the asymptotic relation $(2\pi/L)^3 \sum_{\vec{k}}\to\int d^3k$, $(L/2\pi)^{3/2} A_{\vec{k}}\to A(\vec{k})$ and $(L/2\pi)^{3/2} \varphi_k(t)\to \varphi(k,t)$, where $A(\vec{k})$ and $\varphi(k,t)$ are the annihilation operator and the field mode for continuum $\vec{k}$.}

Consider the expansion in modes of the field operator
\be \label{expansion} \varphi(\vec{x},t )= \sum_{\vec{k}}\left( A_{\vec{k}} \varphi_k(t) + A_{-\vec{k}}^{\dagger} \varphi^*_k(t) \right)  e^{i \vec{k} \vec{x}}\  , \ee 
where $A_{\vec{k}}^{\dagger}$ and $A_{\vec{k}}$ are (time-independent) creation and annihilation operators associated with the modes $\varphi_k(t)$ satisfying the standard commutation relations $[A_{\vec{k}}, A_{\vec{k'}}^{\dagger}]=\delta_{\vec{k},\vec{k}'}$, $[A_{\vec{k}},A_{\vec{k}'}]=0$. 
We study the spectrum of perturbations of the quantum field $\varphi(\vec{x},t)$ as a function of time when the state is described by a density operator $\rho$.  A straightforward calculation shows that\footnote{The mean value of the field, ${\rm Tr}[\rho \varphi(\vec{x},t)]$, vanishes.} 
\be \label{var} {\rm Tr}[\rho \varphi^2(\vec{x},t)] ={\rm Tr}[\rho \sum_{\vec{k}} (1+N_{\vec{k}}+N_{-\vec{k}}) |\varphi_{k}(t)|^2]
=  \sum_{\vec{k}} \big(1+2 {\rm Tr}[\rho N_{\vec{k}}]\big) |\varphi_{k}(t)|^2\ , \ee
where $N_{\vec{k}}\equiv   A^{\dagger}_{\vec{k}}A_{\vec{k}}$ is the number operator, and we have taken into account the definition of $\rho$, and the fact that ${\rm Tr}[\rho]=1$.

A general example, in which the stimulated creation of quanta of the perturbation field is clearly illustrated, is a spacetime for which the cosmological scale factor $a(t)$ has an arbitrary time dependence that asymptotically approaches constant values at early and late cosmic times $t$
\be \label{expfac} a(t)\sim  \left\{\begin{array}{l} a_1 \  \ $as$ \  t\to -\infty \ ,\\  a_2 \  \ $as$ \  t\to +\infty \ . \end{array}\right. \ee
We assume that $a(t)$ is sufficiently smooth and approaches constant values sufficiently fast that the statements we make below are well defined. Examples that allow exact solutions can be found in \cite{parker-toms,parkernature} and, joined smoothly to an arbitrarily long period of de Sitter inflation, in \cite{Glenz-Parker09}. Because at early times the geometry approaches Minkowski spacetime, it is natural to consider in the expansion of the field (\ref{expansion}) normalized field modes such that they approach plane waves at early times,
$\varphi_{k}(t\to -\infty)=e^{-i w_{1 } t}/\sqrt{a_1^3L^3 2 w_{1 }}$, where $t$ and $\vec{x}$ are coordinates of an inertial observer located in the past, and $w_1\equiv |\vec{k}|/a_1$ is the frequency as measured by this observer. The creation and annihilation operators $A_{\vec{k}}^{\dagger}$ and $A_{\vec{k}}$ then create and annihilate quanta representing standard particles in the early time Minkowski spacetime, and ${\rm Tr}[\rho\, N_{\vec{k}}]$ corresponds to the average number of these initial particles.
Then, at early times, expression (\ref{var}) reduces to 
\be {\rm Tr}[\rho\, \varphi^2(\vec{x},t\to - \infty)] =\frac{1}{(2 a^3_1 L^3)} \sum_{\vec{k}} \frac{1}{w_1} (1+2 \,{\rm Tr}[\rho\, N_{\vec{k}}]) \ , \ee
where we can see the contributions coming from vacuum fluctuations and from the quanta that are present in the initial state. 

At late times, as a consequence of the expansion of the universe, the field modes $\varphi_k(t)$ no longer will be purely positive frequency plane-waves. In general, they will be a linear combination of positive and negative frequency plane-waves of the form
\be \label{latetimemode} \varphi_k(t\to+\infty)=\frac{1}{(a^3_2 L^3 2 w_2)^{1/2}} \left( \alpha_k e^{-i w_2 t}+\beta_k e^{i w_2 t} \right) \ , \ee
where $w_2\equiv |\vec{k}|/a_2$ is the frequency of the mode as measured by an inertial observed located in the late time region. The coefficients $\alpha_k$ and $\beta_k$ are the so-called Bogoliubov coefficients (see \cite{parker-toms} for additional details). They depend on the particular form of the expansion factor $a(t)$ and satisfy the relation $ |\alpha_k|^2 -|\beta_k|^2=1$. 

The late-time amplitude of the modes is given by 
\be 
\label{latetime}
|\varphi_k(t\to\infty)|^2=\frac{1}{a^3_2 L^3 2 w_2}  \left[   |\alpha_k|^2 +|\beta_k|^2 +2 \Re{( \alpha_k \beta^*_k e^{-i 2 w_2 t})} \right]\approx \frac{1}{a^3_2 L^3 2 w_2}  (   |\alpha_k|^2 +|\beta_k|^2) \ , \ee
where we have neglected the oscillatory term because it averages to $0$ and is not important for our present discussion. Putting (\ref{latetime}) in (\ref{var}), one obtains\footnote{The analogous equation for the average number of particles present at late times, $Tr[\rho N_k(t \rightarrow \infty) ]$, is derived in \cite{parker68-69-71} [see second paper, Eq. (53)], and in \cite{parkerthesis} [see Appendix CII ]. This stimulated particle creation is confirmed numerically for many e-foldings of inflation in \cite{Glenz-Parker09}.}
\be \label{trace} {\rm Tr}[\rho \varphi^2(\vec{x},t\to\infty)] = \frac{1}{(2 a^3_2 L^3)} \sum_{\vec{k}} \frac{1}{w_2} \bigg[(1+2 |\beta_k|^2)
+2 \ {\rm Tr}[\rho N_{\vec{k}} ] (1+2 |\beta_k|^2) \bigg] \ .\ee

In the previous expression, the first term in square brackets, $(1+2 |\beta_k|^2)$, has two parts, a $1$ coming from the vacuum fluctuations of the late-time Minkowski space, and a $2 |\beta_k|^2$  coming from the spontaneous creation of quanta that would result from the expansion of the universe if there were no quanta present at early-times. The $1$ coming from vacuum fluctuations produces a divergent contribution that is removed by renormalization in the late-time Minkowski space.

The second term in square brackets, $2 \ {\rm Tr}[\rho N_{\vec{k}} ] (1+2 |\beta_k|^2)$, is proportional to the average number of particles in  mode ${\vec{k}}$ that were present in the early-time Minkowski space.  The first part of this term is $2 \ {\rm Tr}[\rho N_{\vec{k}} ]$, which comes from the original particles that were present in the early-time Minkowski space.  If the universe expands by a large factor, $a_2/a_1 \gg 1$, then the density of these original particles is greatly diluted by the expansion, and this term has a negligible effect.  The second part of this term is $4 |\beta_k|^2\ {\rm Tr}[\rho N_{\vec{k}} ]$, which comes from stimulated particle creation.
The presence of $|\beta_k|^2$ in this term tells us that the stimulated creation process is enhanced by the same amplification factor as the spontaneous creation.  Thus, the presence of initial quanta at early times will leave their imprint on the spectrum of perturbations at late times. This is important to take into account in computing the spectrum of primordial perturbations resulting from inflation, as we will show in the remaining sections of this paper.

It is worth noting that renormalization in the late-time Minkowski space does not affect the second term in square brackets in (\ref{trace}), which is proportional to ${\rm Tr}[\rho N_{\vec{k}} ]$.  So the potential impact of renormalization on the two-point function and the power spectrum \cite{Parker07,agulloetal09,agulloetal10PRD}, and on higher order correlation functions do not interfere with the effects of the stimulated creation of perturbations.  Therefore, we will not consider renormalization in the rest of this paper. 

\section{Scalar Power spectrum and stimulated creation of perturbations during inflation  \label{powspec}}

We are now ready to determine the observable effects on the CMB if there were perturbation quanta already present at a time $t_0$ about 60 $e$-foldings before the end of slow-roll inflation. As we showed in the previous section, these quanta would give rise to the stimulated creation of quanta. As we have seen, in the absence of this stimulated creation of quanta, there would be no such observable effects because of the huge dilution of the original quanta during the 60 inflationary $e$-foldings.  As we showed in the last paragraph of section \ref{initialstate} [see (\ref{energydensity1})], the presence of such quanta at time $t_0$ is consistent with the scenario in which the unperturbed part of the inflaton causes inflation through its potential energy density function $V(\phi)$. We loosely refer to $t_0$ as the initial time and the state at time $t_0$ as the initial state. 

Inflation provides an elegant mechanism to generate inhomogeneities in the early universe \cite{inflation2}. These inhomogeneities are believed to have originated as a consequence of the rapidly expanding inflationary geometry. This expansion induces the creation of quantum fluctuations of the inflaton field and also those of metric perturbations. These scalar and tensorial fluctuations evolve with time while their physical wavelength (in momentum space) is smaller than the Hubble radius $R_H\equiv H^{-1}$ during inflation. When their physical wavelength becomes larger than $R_H$ (this is referred to as the exit from the Hubble radius), these fluctuations stop evolving and remain constant. When they re-enter the Hubble radius at a later stage of radiation or matter domination, they constitute the initial conditions for the inhomogeneities in the universe that will further evolve to produce the temperature fluctuations in the CMB and the LSS. We analyze here the production of scalar perturbations during inflation. The extension to tensor perturbations is straightforward.  We consider the general case of single field slow-roll inflation. In this scenario, the homogenous part of the inflaton field, $\phi(t)$, slowly rolls down its potential $V(\phi)$ towards a minimum during the evolution. The slow-roll parameters are defined as $\epsilon=(M_P^2/2)(V'/V)^2$ and $\eta =M_P^2(V''/V)$, where the prime here means derivative with respect to $\phi(t)$ ($M_P=1/\sqrt{8 \pi G}$ is the reduced Planck mass). In the slow-roll regime, defined as $\epsilon, |\eta|\ll 1$, the evolution of the inflaton field is such that the potential $V(\phi(t))$ changes very slowly and takes the role of an effective cosmological constant in the Einstein equations. The universe then expands in a quasi-exponential way with $H\equiv\dot{a}(t)/a(t)\approx\sqrt{(8\pi G/3) V(\phi)}$ remaining nearly constant, where $\dot a \equiv\partial_t a$.  The slow-roll parameters then can be approximated by $\epsilon\approx1/(2 M_P^2) (\dot\phi/H)^2$ and $\eta \approx\ -\ddot\phi/(\dot\phi H)+1/(2 M_P^2) \dot\phi^2/H^2$,.\\

Scalar perturbations are usually characterized by the gauge-invariant quantity $\cal{R}$ (the comoving curvature perturbation) which, during inflation, is given by
\be {\cal{R}}=\Psi+ \frac{H}{\dot\phi}\delta\phi \ , \ee
where $\delta \phi(\vec{x},t)$ represents the inflaton perturbation field and $\Psi$ is a scalar perturbation of the metric, related to the curvature of the spatial metric through $R^{(3)}=4 \nabla^2\Psi/a^2(t)$ (for details see, for instance, \cite{Weinberg2008,baumann}). 
To fix the time and spatial reparametrizations we choose the so-called comoving gauge in which 
\be \label{comgauge} \delta \phi=0\ ,  \hspace{1cm} g_{ij}=a^2(t) [e^{-2{\cal R}} \delta_{ij}+ h_{ij}]\ , \hspace{1cm} \partial_i h_{ij}=h^i_i=0 \ . \ee 
 In this gauge, the inflaton field is unperturbed and all scalar degrees of freedom are encoded in ${\cal R}$. In this section, it is sufficient for us to work to linear order in the perturbation $\cal{R}$.
 
We expand the field in Fourier modes
\be \label{modexp} {\cal R}(\vec{x},t)= \sum_{\vec{k}} \hat {\cal R}_{\vec{k}}(t) e^{i \vec{k} \vec{x}} \ , \hspace{1cm}  \hat{\cal R}_{\vec{k}}(t)=A_{\vec{k}} {\cal R}_k(t)+A^{\dagger}_{-\vec{k}} {\cal R}^*_k(t) \ . \ee 
At linear order, the field ${\cal R}(\vec{x},t)$ is described by the action $S_2=1/2 \int d^3xd\tau z^2 [{\cal{R}}'^2-(\partial {\cal{R}})^2]$, where ${\cal{R}}' \equiv \partial_{\tau} {\cal{R}}$, and the field modes obey the equation \cite{Mukhanov86}
\be \frac{d^2 {\cal R}_k}{d{\tau}^2} + \frac{2}{z}\frac{dz}{d\tau}\frac{d {\cal{R}}_k}{d{\tau}} + k^2{\cal{R}}_k =0 \ , \ee 
where $\tau \equiv \int dt/a(t)$ is the conformal time and $z\equiv a\dot{\phi}/H$. In the slow-roll approximation $z^{-1}dz/d\tau\approx (1+3\epsilon-\eta)/\tau$. The (normalized) solutions to this equation (considering $\epsilon$ and $\eta$ as constants) obeying the adiabatic condition \cite{parker-toms} (and the de Sitter symmetry for $H$ constant) are
\be \label{hankmods} {\cal{R}}_k( \tau) = (-\pi \tau /4L^3 z^2)^{1/2}H^{(1)}_{\mu}(-\tau k) \ , \ee 
where $H^{(1)}_{\mu}$ is a Hankel function with index $\mu= 3/2 + 3\epsilon -\eta$. The amplitude of these modes goes to a constant value a few $e$-folds after the time $\tau_k$ at which the mode $k$ exits from the Hubble radius ($k/a(\tau_k)=H$). It is this fact that makes the variable $\cal{R}$ interesting in the study of scalar perturbations. The spectrum of primordial scalar perturbations generated during inflation is usually characterized by the so-called power spectrum $P_{{\cal R}}$. This  quantity can be defined in terms of the momentum space two-point function evaluated a few $e$-folds after the Hubble exit time $\tau_k$. When the state of the field is given by the vacuum state [defined with respect to the modes (\ref{hankmods})], we have 
\be \langle0| \hat {\cal{R}}_{\vec{k}_1}  \hat  {\cal{R}}_{\vec{k}_2} |0\rangle \equiv \delta_{\vec{k}_1+\vec{k}_2,0} \, P^0_{{\cal R}}(k) \ ,\ee
where\footnote{In the limit $L\to\infty$ this definition is consistent with the fact that the quantity $ \Delta_{\cal R}^2(k)^0$ is the contribution per $d\ln{k}$ to the two-point function at coincident points, $\langle0| {\cal{R}}^2 (\vec{x},t) |0\rangle =\int dk/k \  \Delta_{\cal R}^2(k)^0$.}
\be  \label{delta} P^0_{{\cal R}}(k)=|{\cal R}_k|^2\equiv\left(\frac{2\pi}{L}\right)^3  \frac{\Delta_{\cal R}^2(k)^0 }{4\pi k^3}\ , \hspace{1cm}   \Delta_{\cal R}^2(k)^0=\frac{1}{2M_P^2 \epsilon(\tau_k)}\left(\frac{H(\tau_k)}{2\pi}\right)^2 \ , \ee
where the superscript $0$ indicates that the previous quantities are evaluated using the vacuum state.

Let us suppose now that the state of the field is described by a statistical density operator $\rho$ of the type defined in section \ref{initialstate}. The power spectrum $P^{\rho}_{{\cal R}}(\vec{k})$ in this case is defined by 
\be \label{powrho} {\rm Tr}[ \rho \, \hat {\cal{R}}_{\vec{k}_1}   \hat  {\cal{R}}_{\vec{k}_2} ] \equiv \delta_{\vec{k}_1+\vec{k}_2,0} \, P^{\rho}_{{\cal R}}(\vec{k}) \ ,\ee
where a straightforward computation gives the following expression for $P^{\rho}_{{\cal R}}(\vec{k})$.
\be  \label{newpowspect} P^{\rho}_{{\cal R}}(\vec{k})= P^0_{{\cal R}}(k) (1+{\rm Tr}[\rho N_{\vec{k}}]+ {\rm Tr}[\rho N_{-\vec{k}}]) \ .\ee
Consequently
\be  \label{deltarho}  \Delta_{\cal R}^2(\vec{k})^{\rho} = \Delta^2_{\cal R}(k)^0(1+{\rm Tr}[\rho N_{\vec{k}}]+ {\rm Tr}[\rho N_{-\vec{k}}]) \ ,\ee
where ${\rm Tr}[\rho N_{\vec{k}}]$ is the average number of quanta in the mode $\vec{k}$  present in the initial state and the superscript $\rho$ indicates that a quantity is evaluated using the statistical density operator $\rho$.\footnote{A similar analysis holds for tensor perturbations, and the result is $P^{\rho_t}_{{ h}}(\vec{k})= P^0_{{ h}}(k) (1+{\rm Tr}[\rho_t N^t_{\vec{k}}]+ {\rm Tr}[\rho_t N^t_{-\vec{k}}])$, where $P^{0}_{{ h}}(\vec{k})$ is the power spectrum for tensor perturbations when the initial state of those perturbations is the vacuum state, $\rho_t$ is the statistical density operator describing the initial state of tensor perturbations at the time $t_0$, and $N^t_{\vec{k}}$ is the operator for the number of quanta of tensor perturbations in mode with coordinate momentum ${\vec k}$. As a consequence, the tensor-to-scalar ratio takes the form $r\equiv P^{\rho_t}_{{ h}}/P^{\rho}_{{\cal R}}(\vec{k})=r^0\cdot \alpha$, where $r^0=P^{0}_{{ h}}/P^{0}_{{\cal R}}(\vec{k})=16\epsilon$ and the extra factor $\alpha\equiv (1+{\rm Tr}[\rho_t N^t_{\vec{k}}]+ {\rm Tr}[\rho_t N^t_{-\vec{k}}])/(1+{\rm Tr}[\rho N_{\vec{k}}]+ {\rm Tr}[\rho N_{-\vec{k}}])$ depends on the initial states of tensor and scalar perturbations.} We can observe in the above expression how the stimulated creation of quanta enhances the magnitude of the power spectrum by a factor $ (1+{\rm Tr}[\rho N_{\vec{k}}]+{\rm Tr}[\rho N_{-\vec{k}}])$ and introduces the anisotropies that ${\rm Tr}[\rho N_{\vec{k}}]$ could contain. We consider now the constraints on the quantity ${\rm Tr}[\rho N_{\vec{k}}]$ coming from the present CMB observational data \cite{wmap7}.

First, we note that the high degree of isotropy of the CMB \cite{wmap7} constrains the variation of ${\rm Tr}[\rho N_{\vec{k}}]$ with the direction of $\vec{k}$ to be small. Therefore, it is a good first approximation to suppose that there is no dependence on the direction of ${\vec{k}}$ in the initial state so that ${\rm Tr}[\rho N_{\vec{k}}]\approx {\rm Tr}[\rho N_{-\vec{k}}]$. [However, if the evidence for anisotropies were to become stronger, then it would not be unreasonable to investigate possible anisotropies of the initial state as a potential source for anisotropies and other related anomalies (see, for instance \cite{copietal} and references therein) that may be present in the CMB and LSS)]. In the rest of this paper, we consider only isotropic states of the initial perturbations.

Next, we consider the constraints that come from the measured parameters $A(k_{\rm p})$ and $n_s$ that characterize the observed power spectrum. The observed power spectrum is written in terms of these parameters as follows,

\be  \Delta_{\cal R}^2(k)=A(k_{\rm p}) \left( \frac{k}{k_{\rm p}}\right)^{n_s-1} \ , \ee
where $k_{\rm p}$ is a pivot (or reference) scale. The observations of 7-year WMAP (together with BAO and $H_0$ Mean) \cite{wmap7} provide values for the two parameters,  
\be 
\label{A-n}
A(k_{\rm p})=(2.441^{+0.088}_{-0.092})10^{-9} \ , \hspace{1cm} n_s=0.963\pm0.012 \ , \ee
where $k_{\rm p}=0.002\  {\rm Mpc}^{-1}$. 

Our theoretical predictions now follow from (\ref{deltarho}).
Using (\ref{delta}) and (\ref{deltarho}), one finds that
\be \label{mag} A^{\rho}(k_{\rm p})= \frac{1}{2 M_P^2 \epsilon(\tau_{k_{\rm p}})}\left(\frac{H(\tau_{k_{\rm p}})}{2\pi}\right)^2 (1+2 \,{\rm Tr}[\rho N_{\vec{k}_{\rm p}}])  \ , \ee
and 
\be \label{index} n^{\rho}_s\equiv \frac{d\ln{\Delta_{\cal R}^2(k)^{\rho}}}{d\ln{k}}+1= 1 -6\epsilon + 2\eta
+ \frac{d\ln{ (1+2\, {\rm Tr}[\rho N_{\vec{k}}])}}{d \ln k} \ . \ee
We now impose the observational constraints that
\be
\label{Arho}
A^{\rho}(k_{\rm p}) = A(k_{\rm p}) = (2.441^{+0.088}_{-0.092})10^{-9}, \ \ \ \ \  n^{\rho}_s = n_s = 0.963\pm0.012\ ,
\ee
with the measured values in (\ref{A-n}). 

Because the value of $H^2/\epsilon$  is uncertain at the time $\tau_{k_{\rm p}}$ when the mode at the pivot scale exited the Hubble radius, the first condition in (\ref{Arho}) does not strongly constrain the value of ${\rm Tr}[\rho N_{\vec{k}_{\rm p}}]$. On the other hand, the second condition in (\ref{Arho}) imposes more severe restrictions on the dependence of ${\rm Tr}[\rho N_{\vec{k}}]$ on $k$.  In particular, this condition constrains the quantity $\left| \frac{d\ln{ (1+2{\rm Tr}[\rho N_{\vec{k}}])}}{d \ln k}\right| $ to be not much bigger than the slow-roll parameters $\epsilon \sim |\eta|\sim 1/100$.

\section{Primordial non-gaussianities from stimulated creation of perturbation during inflation \label{bispec}}

As we have seen, the power spectrum alone has limited potential in revealing detailed information about the dynamics and interactions of the field or fields driving inflation. Higher-order correlation functions, such as the three-point function, are more sensitive to those aspects of inflation (see for instance \cite{whitepaper}). In the following, we compute the three-point function in momentum space, to leading order in the perturbations, when the initial state of scalar perturbations is given by a density operator $\rho$ of the type introduced in section \ref{initialstate}. We obtain the values of the primordial non-gaussianites, those that were present after inflation ended, before their reentry into the Hubble horizon of the radiation- or matter-dominated universe.

The full computation of the three-point function using the vacuum state was given by Maldacena in \cite{maldacena}. Here, we generalize those computations to the case of a more general state described by a statistical density operator of the type discussed in section \ref{initialstate}. As explained there, this density operator allows for the possibility that there were perturbations already present at a time $t_0$, about 60 $e$-foldings before the end of inflation. Our calculation includes the effects of the stimulated creation of perturbation quanta. This stimulated creation is the reason why the presence of such quanta early in the inflationary epoch cannot be ignored. We limit our calculation to the case of single-field inflation. 

Let us consider the three point function in momentum space for comoving curvature perturbation ${\cal{R}}(x)$, parametrized in the following way 
\be  {\rm Tr}[\rho \hat{\cal{R}}_{\vec{k}_1}(\tau)\hat{\cal{R}}_{\vec{k}_2}(\tau)\hat{\cal{R}}_{\vec{k}_3}(\tau)] = 
\delta_{({\tiny\sum} {\vec{k}_i}),0} 
 B_{{\cal{R}}}(\vec{k}_1,\vec{k}_2,\vec{k}_3) \ , \ee
where the Kronecker delta is a consequence of translational invariance and enforces the condition that ${\sum} {\vec{k}}_i=0$. The quantity $B_{{\cal{R}}}(\vec{k}_1,\vec{k}_2,\vec{k}_3)$ is known as the bispectrum. As in the previous section, we work in the comoving gauge (\ref{comgauge}).
In order to compute the bispectrum at leading order in the perturbations, we expand the inflationary action around the spatially homogenous solution to third order
\be S=S_0(\phi,g_{\mu\nu})+S_2({\cal{R}}^2)+S_3({\cal{R}}^3)+... \ee
The background action $S_0(\phi,g_{\mu\nu})$, which depends on the background inflaton field $\phi(t)$ and the background metric, defines the Hubble rate $H$ and the slow-roll parameters. The action 
$S_2=1/2 \int d^3xd\tau\, z^2 [{\cal{R}}'^2-(\partial {\cal{R}})^2]$, where ${\cal{R}}' \equiv \partial_{\tau} {\cal{R}}$, defines the free evolution of the mode functions ${\cal{R}}_k(\tau)$. The action $S_3$ defines the interaction Hamiltonian. The form of this action was computed in \cite{maldacena}. We refer the reader to \cite{maldacena} for further details. Maldacena obtained a simple form of the cubic terms in the action by using the following field redefinition 
\be {\cal{R}}={\cal{R}}_c+\left[\frac{1}{2} \frac{\ddot{\phi}}{\dot{\phi} H} +\frac{1}{8 M_P^2} \frac{\dot{\phi}^2}{H^2}\right] {\cal{R}}^2_c+\frac{1}{4 M_P^2}  \frac{\dot{\phi}^2}{H^2} \partial^{-2} ({\cal{R}}_c\partial^2 {\cal{R}}_c) \ . \ee
At the lowest order in the perturbations, ${\cal{R}}_c$ is the same as ${\cal{R}}$, so at that order the mode functions of ${\cal{R}}_c$ and ${\cal{R}}$ are also the same. At leading order in the slow-roll parameters, $S_3=-\int d\tau H_{\rm int}$, where 
\be H_{\rm int}(\tau)=-\int d^3x \ a^3(\tau) \left(\frac{\dot\phi}{H}\right)^4 H M_P^{-2} \ {\cal{R}}'^2_c\partial^{-2}{\cal{R}}'_c \ . \ee
We compute the three-point function at leading order in curvature perturbations and in the slow-roll parameters by using time-dependent perturbation theory. In the interaction picture, the three-point function in momentum space is given by
\bea & & \hspace{-1cm}{\rm Tr}[\rho \hat{\cal{R}}_{\vec{k}_1}(\tau)\hat{\cal{R}}_{\vec{k}_2}(\tau)\hat{\cal{R}}_{\vec{k}_3}(\tau)] =  \\ &=& {\rm Tr} \left[\rho \left( \bar{T} e^{i\int_{\tau_0}^{\tau} H^I_{\rm int}(\tau') d\tau'}\right) \hat{\cal{R}}^I_{\vec{k}_1}(\tau)\hat{\cal{R}}^I_{\vec{k}_2}(\tau)\hat{\cal{R}}^I_{\vec{k}_3}(\tau)   \left( Te^{-i\int_{\tau_0}^{\tau} H^I_{\rm int}(\tau') d\tau'}\right) \right]= \nonumber \\ &=& {\rm Tr}[\rho \hat{\cal{R}}^I_{\vec{k}_1}(\tau)\hat{\cal{R}}^I_{\vec{k}_2}(\tau)\hat{\cal{R}}^I_{\vec{k}_3}(\tau)] -i \int_{\tau_0}^{\tau} d\tau' \, {\rm Tr}\big[\rho [  \hat{\cal{R}}^I_{\vec{k}_1}(\tau)\hat{\cal{R}}^I_{\vec{k}_2}(\tau)\hat{\cal{R}}^I_{\vec{k}_3}(\tau) ,H^I_{\rm int}(\tau')] \big] + {\cal{O}}(H^2_{\rm int}) \nonumber \ ,\eea
where $T$ and $\bar{T}$ indicate the time and anti-time ordered product respectively,  the superscript $I$ indicates operators in the interaction picture, and $\tau_0$ is the conformal time corresponding to the cosmic proper time $t_0$ at about $60$ $e$-foldings before the end of inflation, as discussed earlier. Because comoving curvature perturbations $\hat{\cal{R}}^I_{\vec{k}}$ remain constant outside the horizon, we can evaluate the previous quantity at late-times, when $\tau\to0$. Additionally, in the computation we assume that $\vec{k}_i\neq 0$. The two terms in the last line of the previous equation can be computed explicitly (see Appendix for details) and the result is
\bea & &\hspace{- 1cm} {\rm Tr}[\rho \hat{\cal{R}}^I_{\vec{k}_1}(0)\hat{\cal{R}}^I_{\vec{k}_2}(0)\hat{\cal{R}}^I_{\vec{k}_3}(0)]=\\ & =& \nonumber \delta_{({\tiny\sum} {\vec{k}_i}),0}P^0_{\cal{R}}(k_1) P^0_{\cal{R}}(k_2) \left[ \frac{\ddot{\phi}}{\dot{\phi} H} + \frac{1}{4 M_P^2}\frac{\dot{\phi}^2}{H^2} \left(1+\frac{k_1^2+k_2^2}{k_3^2}\right)\right] F(\rho, \vec{k}_1 , \vec{k}_2)+{\rm cyclic\ permut} \ , \eea
\bea  & &\hspace{- 2.5cm}  -i  \int_{\tau_0}^{0} d\tau' \, {\rm Tr}\big[\rho [  \hat{\cal{R}}^I_{\vec{k}_1}(0)\hat{\cal{R}}^I_{\vec{k}_2}(0)\hat{\cal{R}}^I_{\vec{k}_3}(0) ,H^I_{\rm int}(\tau')] \big]=\\ \nonumber &=& \delta_{({\tiny\sum} {\vec{k}_i}),0} P^0_{\cal{R}}(k_1) P^0_{\cal{R}}(k_2)\frac{2}{M_P^2}\frac{\dot{\phi}^2}{H^2} \frac{k_1^2 k_2^2}{k_3^3} G(\rho, \vec{k}_1,\vec{k}_2,\vec{k}_3) +{\rm cyclic\ permut} \ , \eea
where
\be F(\rho, \vec{k}_1 , \vec{k}_2)={\rm Tr}[\rho (2 N_{\vec{k}_1}+1) (2 N_{\vec{k}_2}+1)]-\delta_ {\vec{k}_1,\vec{k}_2} {\rm Tr}[\rho N_{\vec{k}_1}(N_{\vec{k}_1}+1)]\ ,\ee
and $G(\rho, \vec{k}_1,\vec{k}_2,\vec{k}_3)$ is a symmetric function under cyclic permutations of $( \vec{k}_1,\vec{k}_2,\vec{k}_3)$ given by
\bea& &  \hspace{-1.5cm} G(\rho, \vec{k}_1,\vec{k}_2,\vec{k}_3)=\\ \nonumber &=&\frac{(1-\cos{( k_t \tau_0)})}{k_t} \bigg[ \bigg({\rm Tr}[\rho N_{\vec{k}_1} N_{\vec{k}_2}]+{\rm Tr}[\rho N_{\vec{k}_1}]+{\rm cyclic\ permut} \bigg)+1- \\ \nonumber &-&\left( \frac{\delta_{\vec{k}_1,\vec{k}_2}}{2}   {\rm Tr}[\rho N_{\vec{k}_1} (N_{\vec{k}_1} +1)] +{\rm cyclic\ permut} \right)  \bigg] +\\ \nonumber  &+ & \bigg[ \frac{(1-\cos{( \tilde{k}_1 \tau_0)})}{\tilde{k}_1} \bigg({\rm Tr}[\rho N_{\vec{k}_1} N_{\vec{k}_2}]+{\rm Tr}[\rho N_{\vec{k}_1} N_{\vec{k}_3}]-{\rm Tr}[\rho N_{\vec{k}_2} N_{\vec{k}_3}]+{\rm Tr}[\rho N_{\vec{k}_1}] +\\ \nonumber &+ & \frac{\delta_{\vec{k}_2,\vec{k}_3}}{2}   {\rm Tr}[\rho N_{\vec{k}_2} (N_{\vec{k}_2}+1)] \bigg) +{\rm cyclic\ permut}  \bigg]  \eea
where $k_t\equiv k_1+k_2+k_3$ and $\tilde{k}_i\equiv k_t-2 k_i$. As we mentioned at the end of section \ref{powspec}, we have assumed that ${\rm Tr}[\rho N_{\vec{k}}]\approx {\rm Tr}[\rho N_{-\vec{k}}]$ in the previous computations. (If it were necessary, it would be easy to introduce the effects of anisotropies.) In the previous expressions, the quantities $H$, $\dot\phi$ and $\ddot\phi$ 
appear in combinations that can be expressed in terms of the slow-roll parameters, which can be regarded as nearly constant during inflation.  Putting both terms together and taking into account (\ref{delta}) and (\ref{newpowspect}), we obtain 
\bea \label{threepoint} & &\hspace{-1cm} {\rm Tr}[\rho \hat{{\cal{R}}}_{\vec{k}_1}(0)\hat{{\cal{R}}}_{\vec{k}_2}(0)\hat{{\cal{R}}}_{\vec{k}_3}(0)]=\delta_{({\tiny\sum} {\vec{k}_i}),0} \  P^{\rho}_{\cal{R}}(\vec{k}_1) P^{\rho}_{\cal{R}}(\vec{k}_2)\times \\ \nonumber &  & \left[\frac{1}{2} \left(3\epsilon-2\eta+\epsilon \frac{k_1^2+k_2^2}{k_3^2}\right) f(\rho, \vec{k}_1 , \vec{k}_2)+4\epsilon \frac{k_1^2 k_2^2}{k_3^3} \,g(\rho, \vec{k}_1 , \vec{k}_2,\vec{k}_3) \right] + {\rm cyclic\  permut}  \ , \eea
where we have defined
 \be  f(\rho, \vec{k}_1 , \vec{k}_2)\equiv  \frac{F(\rho, \vec{k}_1 , \vec{k}_2)}{{\rm Tr}[\rho (2 N_{\vec{k}_1}+1)]{\rm Tr}[\rho (2 N_{\vec{k}_2}+1) ]} \ , \ee
\be g(\rho, \vec{k}_1 , \vec{k}_2,\vec{k}_3)\equiv \frac{G(\rho, \vec{k}_1 , \vec{k}_2,\vec{k}_3)}{{\rm Tr}[\rho (2 N_{\vec{k}_1}+1)]{\rm Tr}[\rho (2 N_{\vec{k}_2}+1) ]} \ . \ee
The denominators in $f$ and $g$ arise from the definition of $P^{\rho}_{\cal{R}}(\vec{k})$. Expression (\ref{threepoint}) constitutes our result for the bispectrum when the initial state is given by the density operator $\rho$. 

If there are no particles (i.e., quanta of the perturbation field) present at time $\tau_0$, then ${\rm Tr}[\rho N_{\vec{k}}] = 0$ for all $\vec{k}$, in which case we have $f = 1$ and $g = (1-\cos{( k_t \tau_0)})/k_t$. 
We can choose $\tau_0$ such that at that time all the modes that we observe in the CMB are well inside the Hubble radius, so
$k_i \tau_0=k_i/(a(\tau_0)H)\gg1$ and hence $|k_t \tau_0|\gg 1$.  Averaging over the large range of uncertainty in this value, it is reasonable for us to take $1 - \cos{( k_t \tau_0)} \simeq 1$. Then (\ref{threepoint}) agrees with the result obtained in \cite{maldacena}, in which the pure vacuum state was considered.

The presence of particles in the initial state introduces new interesting features. The most remarkable new element is the presence of the denominators proportional to $\tilde{k}_i$ present in the function $G(\rho, \vec{k}_1,\vec{k}_2,\vec{k}_3)$. The fact that $\tilde{k}_i$ can approach arbitrarily close to zero in some configurations opens the possibility of having a large contribution coming from those configurations, producing an enhancement of the non-gaussianities. This fact was already noticed for a pure initial state given by a Bogoliubov transformation of the Bunch-Davies vacuum in \cite{holman}\cite{meerburg}. There, the momentum configurations corresponding to flattened triangles for which two momenta are collinear were considered. However, as we will show, the most pronounced enhancement appears for configuration in which one of the three momenta is much smaller than the other two, the so called squeezed configuration.  In the following, we particularize expression (\ref{threepoint}) for each of the flattened and squeezed configurations, respectively.

\begin{itemize} 

\item{\bf Flattened configuration}
\end{itemize} 

Let us consider first the case of flattened configurations, in which the momenta forming the triangle are such that 
$\vec{k}_2\approx \vec{k}_3\approx -\vec{k}_1/2$. For this configuration we have $k_t\approx 4 k_2$, $\tilde{k}_1\approx 0$ and $\tilde{k}_2\approx\tilde{k}_3\approx 2 k_2$. 

In the case with no particles in the initial state, we have $f(\rho, \vec{k}_i , \vec{k}_{j\neq i})=1$ and $g(\rho, \vec{k}_1,\vec{k}_2,\vec{k}_3)=(1-\cos{(k_t \tau_0)})/k_t\approx 1/k_t$, where, as before, we have averaged the cosine to zero. Equation (\ref{threepoint}) reduces then to
\be\label{f0} \langle 0|\hat{{\cal{R}}}_{\vec{k}_1}(0)\hat{{\cal{R}}}_{\vec{k}_2}(0)\hat{{\cal{R}}}_{\vec{k}_3}(0)|0\rangle \approx\delta_{({\tiny\sum} {\vec{k}_i}),0}  \  P^{0}_{\cal{R}}(k_1) P^{0}_{\cal{R}}(k_2)\   (31 \epsilon-10\eta) \ .\ee

Let us consider now the case in which the initial state is described by the density operator $\rho$. We can estimate the values of $f$ and $g$ by assuming that  there are no significant correlations between initial particles with different momentum, so we can approximate\footnote{For the density operator of the form given in (\ref{rho1}), this assumption implies that $p(n_1(\vec{k}_1),n_2(\vec{k}_2),...)\approx p(n_1(\vec{k}_1))p(n_2(\vec{k}_2))...\, $.} ${\rm Tr}[\rho N_{\vec{k}}N_{\vec{k'}}]\sim {\rm Tr}[\rho N_{\vec{k}}] {\rm Tr}[\rho N_{\vec{k'}}]$. Additionally, if we assume that the average number of initial particles is equal to or greater than one, 
and as we showed earlier, that the value of ${\rm Tr}[\rho N_{\vec{k}}]$ does not vary significantly with $\vec{k}$, it follows that $f$ is the ratio of quantities of order $({\rm Tr}[\rho N_{\vec{k}}])^2$ and that $f(\rho, \vec{k}_i , \vec{k}_{j\neq i})\sim {\cal{O}}(1)$.

On the other hand, because in this configuration $\tilde{k}_1\approx 0$, the quantity $g$ is essentially given by the term proportional to $1/\tilde{k}_1$, which has the form
 \bea g(\rho, \vec{k}_1,\vec{k}_2,\vec{k}_3)&\approx& \frac{1-\cos{(\tilde{k}_1 \tau_0)}}{ \tilde{k}_1}\  \frac{ 2 {\rm Tr}[\rho N_{\vec{k}_1} N_{\vec{k}_2}]-\frac{1}{2} {\rm Tr}[\rho N_{\vec{k}_2}^2]+ {\rm Tr}[\rho N_{\vec{k}_1}]+\frac{1}{2} {\rm Tr}[\rho N_{\vec{k}_2}]}{{\rm Tr}[\rho (2 N_{\vec{k}_1}+1)]{\rm Tr}[\rho (2 N_{\vec{k}_2}+1) ]} \nonumber \\ &\approx& \frac{\tilde{k}_1^2 \tau_0^2}{2! \tilde{k}_1}  \times {\cal O}(1) \ ,\eea
where we have expanded $(1-\cos{(\tilde{k}_1 \tau_0))}$ for $|\tilde{k}_1 \tau_0|\to 0$ and have again used the fact that the factor involving the average number of initial particles is of order 1. Note the extra factor $\tilde{k}_1^2 \tau_0^2$ introduced by the expansion of the cosine. Expression (\ref{threepoint}) then reduces to
\be \label{frho} {\rm Tr}[\rho \hat{{\cal{R}}}_{\vec{k}_1}(0)\hat{{\cal{R}}}_{\vec{k}_2}(0)\hat{{\cal{R}}}_{\vec{k}_3}(0)] \approx\delta_{({\tiny\sum} {\vec{k}_i}),0}  \  P^{\rho}_{\cal{R}}(\vec{k}_1) P^{\rho}_{\cal{R}}(\vec{k}_2) \big[ {\cal O}( \epsilon, \eta)  + 18 \epsilon (k_2 \tau_0) (\tilde{k}_1 \tau_0) \big] \ .\ee

If we write the magnitude of the three-point function in the flattened configuration  in terms of a parameter $\ f_{NL}$ in the form
\be \langle \hat{{\cal{R}}}_{\vec{k}_1}(0)\hat{{\cal{R}}}_{\vec{k}_2}(0)\hat{{\cal{R}}}_{\vec{k}_3}(0)\rangle=\delta_{({\tiny\sum} {\vec{k}_i}),0}  \, P_{\cal{R}}(\vec{k}_1) P_{\cal{R}}(\vec{k}_2) \ f_{NL}\ ,\ee
we obtain from (\ref{f0}) and (\ref{frho}) the following ratio of the value $f^{\rho}_{NL}$ for an initial state described by the density operator $\rho$, to the value $f^{0}_{NL}$ for the vacuum initial state
\be \frac{f_{NL}^{\rho}}{f_{NL}^{0}}\sim  \frac{18 \epsilon}{31 \epsilon-10\eta} (k_2 \tau_0) (\tilde{k}_1 \tau_0) + {\cal{O}}(1) \sim (k_2 \tau_0) (\tilde{k}_1 \tau_0)+ {\cal{O}}(1)    \ . \ee
Since $k_2$ is well inside the horizon at  the early time $\tau_0$, we have that $|k_2\tau_0|\gg1$. On the other hand, we have that $\tilde{k}_1\tau_0\to 0$ for the flattened configuration. We can conclude that there is not necessarily a large enhancement in the primordial non-gaussianities for flattened configurations when quanta are present in the initial state for the canonical interaction lagrangian.  As shown in \cite{holman}, different conclusions could be reached if higher-derivative interactions were considered.

\begin{itemize} 

\item{\bf Squeezed configuration}

\end{itemize}

Let us consider now the configuration in which two of the momenta forming the triangle are much bigger than the third one, for instance $k_1\approx k_2\gg k_3$. Then, we have $k_t\approx 2 k_1$, $\tilde{k}_1\approx \tilde{k}_2\approx k_3$, $\tilde{k_3}\approx 2 k_1$. 
The three-point function in the squeezed configuration is often parametrized in the following way \cite{komatsuspergel, komatsuthesis}
\be \langle \hat{{\cal{R}}}_{\vec{k}_1}(0)\hat{{\cal{R}}}_{\vec{k}_2}(0)\hat{{\cal{R}}}_{\vec{k}_3}(0)\rangle=\delta_{({\tiny\sum} {\vec{k}_i}),0} \  P_{\cal{R}}(\vec{k}_1) P_{\cal{R}}(\vec{k}_3) \frac{12}{5} f_{NL} \ . \ee
In the case with no particles in the initial state, we have $f(\rho, \vec{k}_i , \vec{k}_{j\neq i})=1$ and $g(\rho, \vec{k}_1,\vec{k}_2,\vec{k}_3)=(1-\cos{(k_t \tau_0)})/k_t\simeq 1/k_t$. Then, expression (\ref{threepoint}) reduces to
\be\langle 0|\hat{{\cal{R}}}_{\vec{k}_1}(0)\hat{{\cal{R}}}_{\vec{k}_2}(0)\hat{{\cal{R}}}_{\vec{k}_3}(0)|0\rangle \approx\delta_{({\tiny\sum} {\vec{k}_i}),0}  \  P^{0}_{\cal{R}}(k_1) P^{0}_{\cal{R}}(k_3)\  (6 \epsilon-2\eta)=\delta_{({\tiny\sum} {\vec{k}_i}),0} \  P^{0}_{\cal{R}}(k_1) P^{0}_{\cal{R}}(k_3)\  (1-n^0_s)\ ,\ee
where the scalar spectral index in the vacuum case is $n^0_s = 1-6 \epsilon+2\eta$.
Therefore, we have $f_{NL}^0=\frac{5}{12} (1-n_s)$, in agreement with previous analysis \cite{maldacena}.\\

In the case where the initial state is described by the density operator $\rho$, we still have that  $f(\rho, \vec{k}_i , \vec{k}_{j\neq i})\sim1$, provided the initial average number of quanta in the modes $\vec{k}_i$ and $\vec{k}_j$ is equal to or greater than one, and correlations between particles with different momentum are not large. Additionally, we have 
\be g(\rho, \vec{k}_1,\vec{k}_2,\vec{k}_3)\approx \frac{1-\cos{(k_3 \tau_0)}}{k_3}\  \frac{ 2 {\rm Tr}[\rho N_{\vec{k}_1} N_{\vec{k}_2}]+{\rm Tr}[\rho N_{\vec{k}_1}]+{\rm Tr}[\rho N_{\vec{k}_2}]}{{\rm Tr}[\rho (2 N_{\vec{k}_1}+1)]{\rm Tr}[\rho (2 N_{\vec{k}_2}+1) ]} \sim \frac{1-\cos{(k_3 \tau_0)}}{k_3} \times {\cal O}(1)\ . \nonumber \ee
It is important to notice that, because $\vec{k}_3$ corresponds to a mode whose effects we wish to measure in the present CMB or LSS, its wavelength must have been well inside the Hubble radius at the early time $\tau_0$. That implies that $k_3 \tau_0=k_3/(a(\tau_0) H) \gg 1$, where $H$ is the inflationary Hubble rate. Therefore, we can not expand the $\cos{(k_3 \tau_0)}$ above, and we have
\be {\rm Tr}[\rho \hat{{\cal{R}}}_{\vec{k}_1}(0)\hat{{\cal{R}}}_{\vec{k}_2}(0)\hat{{\cal{R}}}_{\vec{k}_3}(0)] \approx\delta_{({\tiny\sum} {\vec{k}_i}),0} \  P^{\rho}_{\cal{R}}(\vec{k}_1) P^{\rho}_{\cal{R}}(\vec{k}_3) \, 4 \, \epsilon \,\frac{k_1}{k_3} \, (1-\cos{(k_3 \tau_0)})  \ ,\ee
from which we can obtain
\be f_{NL}^{\rho}\approx  \frac{20}{12} \epsilon \frac{k_1}{k_3} (1-\cos{(k_3 \tau_0)})\approx \frac{5}{3} \,\epsilon \,\frac{k_1}{k_3} \ ,\ee
where we have again averaged the cosine of large argument to zero. Note that $f_{NL}^{\rho}$ is proportional to the slow-roll parameter $\epsilon$, so the size of the non-gaussianities is exactly zero for an exact de Sitter inflation. We can now compute, for the squeezed configuration, the ratio $f_{NL}^{\rho}/f_{NL}^{0}$ between the magnitude of the non-gaussianities when there are particles present in the initial state and their magnitude in the case of the initial vacuum state
\be \label{ratio} \frac{f_{NL}^{\rho}}{f_{NL}^{0}}\approx \frac{k_1}{k_3} \frac{4 \epsilon}{6\epsilon -2 \eta} \approx  \frac{k_1}{k_3}  \ . \ee
Because in this squeezed configuration we have $k_1\gg k_3$, it follows that  ${f_{NL}^{\rho}}/{f_{NL}^{0}}\gg 1$ . 

We can estimate the value of the ratio in (\ref{ratio}) by considering the range of scales that we can measure in the CMB. The range of angular multipoles $\ell$ for which we have some confidence about the measurements of the temperature fluctuations of the CMB (i.e., for which uncertainities coming from cosmic variance, Sunyaev-Zeldovich effect, etc., can be neglected) is approximately $\ell\in (10, 2000)$. On the other hand, for $\ell\gg 1$ the contribution of comoving momenta $k$ to the multipole $\ell$ is peaked arround $k\approx \ell/ r_L$, where $r_L$ is the radial coordinate of the surface of last scattering.\footnote{This is just a consequence of the properties of the spherical Bessel functions $j_{\ell}(k r_L)$ appearing in the relation between $k$-Fourier space and angular-Fourier space. For further details see, for instance, \cite{Weinberg2008}. } Therefore, the maximum enhancement is achieved when $k_1\approx 2000/r_L$ and $k_3\approx 10/r_L$. In that case $ \frac{f_{NL}^{\rho}}{f_{NL}^{0}}\sim 200$. It is important to note that this enhancement takes place when the number of initial quanta is equal to or greater than one, but does not depend on the exact value of the average number of initial quanta. This estimate is within the range permitted by the effects of backreaction, as discussed after (\ref{energydensity1}).

The analysis of this section shows that, when perturbation quanta are present in the initial state at the onset of inflation (or at an early time when perturbation wavelengths we can observe today were within the inflationary Hubble radius), the size of non-gaussianties produced during inflation would be considerably enhanced. The strongest enhancement would occur when the momenta appearing in the bispectrum form a squeezed configuration, for which there could be an enhancement (as compared to the initial vacuum state) of order one hundred in the parameter $f_{NL}$ that is usually employed to characterize non-gaussianities.

It is important to mention here that our result contrasts with those reached in some previous analyses of non-gaussianities. In \cite{creminellizaldarriaga}, they proposed  a consistency relation for the three-point function of curvature perturbations in momentum space in the squeezed limit. Their consistency relation asserts that the size of the bispectrum in the squeezed limit is proportional to $(1 - n_s)$, with the only assumption being that the inflaton is the only dynamical field. It follows from their consistency relation that, for those single-field inflationary models, the size of the parameter $f_{NL}$ that characterizes the primordial non-gaussianities in the squeezed limit must be of order of the slow-roll parameters. Any detection of non-gaussianities for the squeezed configuration would then rule-out  any inflationary scenario with a single dynamical field. Recently, some weak aspects of the argument used in \cite{creminellizaldarriaga} were pointed out in \cite{ganckomatsu}, where a more rigorous proof of the consistency relation was attempted. However, in \cite{ganckomatsu} they pointed out that they were not able to derive the consistency relation in all generality. They did succeed in proving it more rigorously for the vacuum initial state, but not for the general class of states we have considered in the present paper.
Our results show that when particles are present in the initial state, there are new contributions to the three-point function, or bispectrum, that come from perturbative interactions among the quanta produced by stimulated creation resulting from the presence of initial particles. These contributions can be as much as two orders of magnitude larger than the contribution coming from the vacuum.

\section{Conclusions}

In this paper, we have analyzed the primordial spectrum of scalar perturbations produced during cosmic inflation when the initial quantum state is not necessarily  the vacuum state. We have restricted our analysis to the scenario of single-field slow-roll inflation with the canonical Lagrangian. We have considered an initial state described by a statistical density operator $\rho$. In this mixed state, there are quanta of the perturbation field (i.e., particles) already present at the time $t_0$ when the longest wavelength modes that we can observe today exited the Hubble radius during inflation. If this time is around 60 $e$-folds before the end of inflation, we do not need to consider trans-planckian excitations in  $\rho$. The density operator $\rho$ is formed from a mixture of pure states having definite numbers of perturbation quanta present at $t_0$.  This density operator is natural in models of inflation in which our observable universe arises from the inflation of a small patch of a larger universe. 

Theoretical constraints can be imposed on the maximum average numbers of initial quanta, ${\rm Tr}[\rho N_{\vec{k}}]$, that are present in $\rho$ for the occupied modes $\vec{k}$. These constraints can be obtained by imposing renormalizability of the theory and demanding that backreaction effects from the initial quanta do not modify the background inflationary dynamics. These conditions restrict the average number of initial quanta but, as shown at the end of section \ref{initialstate}, do not necessarily restrict ${\rm Tr}[\rho N_{\vec{k}}]$ to be less than one. 
 
When  $\rho$ contains a non-zero average initial number of quanta, the phenomenon of stimulated creation of perturbations occurs during inflation. We have computed the expression for the power spectrum and the bispectrum of scalar perturbations by taking into account the stimulated creation process. We have shown that the quanta present in the initial state do leave an imprint in the power spectrum and bispectrum of perturbations. Hence, present and future observations of the CMB and LSS can be used to constrain the initial state. In section \ref{powspec}, we showed how the expression for the power spectrum is modified by the presence of initial quanta. We used the observational data for the amplitude of the power spectrum and the spectral index to constrain the function ${\rm Tr}[\rho N_{\vec{k}}]$. The isotropy of the CMB and the value of the spectral index severely restrict the variation of ${\rm Tr}[\rho N_{\vec{k}}]$ with $\vec{k}$. However, due to our uncertainty in the value of the $\epsilon$ slow-roll parameter and the Hubble rate during inflation appearig in the expression for the amplitude of the power spectrum, no significant constraints can be obtained on the magnitude of ${\rm Tr}[\rho N_{\vec{k}}]$ from the power spectrum.

In contrast, the bispectrum (the three-point function in momentum space) is much more sensitive to the size of ${\rm Tr}[\rho N_{\vec{k}}]$ and constitutes a more suitable tool to probe the initial state. We have obtained, in section \ref{bispec}, the complete expression for the bispectrum when the initial state is described by $\rho$ [expression (\ref{threepoint})], which includes the vacuum state as a special case. The most remarkable effect of the initial quanta is an enhancement of the bispectrum for certain configurations of the three different momenta involved. We have found that the largest enhancement appears for configurations in which two of the momenta are much bigger than the third one (for instance, $k_1\approx k_2\gg k_3$), the so-called squeezed configuration. In this limit, for the range of perturbations accessible to observations, the size of the parameter $f_{NL}$ parametrizing the non-gaussianities can be as much as two orders of magnitude larger than its value for the pure vacuum initial state.

This result is important for interpreting the observations made by the PLANCK satellite and by surveys of large-scale structure to determine the size of non-gaussianities.  Our results indicate that, if non-gaussianities are found to be observable in the squeezed limit, then single-field inflation is not necessarily ruled out, as would follow from \cite{creminellizaldarriaga}. Instead, such an observation could be interpreted as a consequence of a non-vacuum initial state, and valuable information about this state could then be obtained from those observations. \\

\noindent { \bf Acknowledgements.} We thank J. Navarro-Salas and G. J. Olmo for useful comments and suggestions. This
work have been partly supported by an NSF grant and by
a University of Wisconsin-Milwaukee RGI grant, and by the Spanish grants FIS2008-06078-C03-02 and  the  Consolider-Ingenio 2010 Programme CPAN (CSD2007-00042).

\section*{Appendix: Explicit computations of the three-point function in momentum space}

In this appendix we present some details of the computation of the three-point function used in section \ref{bispec}. Let us consider that the initial state of scalar perturbations is described by the density operator $\rho$ introduced in section \ref{initialstate}. Additionally, as we mentioned at the end of section \ref{powspec}, the isotropy of the CMB constrains the anisotropies contained in $\rho$, so we assumed that ${\rm Tr}[\rho N_{\vec{k}}]\approx {\rm Tr}[\rho N_{-\vec{k}}]$. The generalization to the non-isotropic case is straightforward. Following Maldacena \cite{maldacena}, we employ the following field redefinition 
\be  \label{filedred} {\cal{R}}={\cal{R}}_c+\left[\frac{1}{2} \frac{\ddot{\phi}}{\dot{\phi} H} +\frac{1}{8 M_P^2} \frac{\dot{\phi}^2}{H^2}\right] {\cal{R}}^2_c+\frac{1}{4 M_P^2}  \frac{\dot{\phi}^2}{H^2} \partial^{-2} ({\cal{R}}_c\partial^2 {\cal{R}}_c) \ . \ee
Using this field redefinition, the interaction hamiltonian takes the simple form
\be H_{\rm int}(\tau)=-\int d^3x \ a(\tau)^3 \left(\frac{\dot\phi}{H}\right)^4 H M_P^{-2} \ {\cal{R}}'^2_c\partial^{-2}{\cal{R}}'_c \ . \ee
In order to compute the three-point function in momentum space, we use time dependent perturbation theory at leading order in perturbations. Then we have
\bea \label{pert} & & {\rm Tr}[\rho \hat{\cal{R}}_{\vec{k}_1}(\tau)\hat{\cal{R}}_{\vec{k}_2}(\tau)\hat{\cal{R}}_{\vec{k}_3}(\tau)] =  \\ &=& {\rm Tr }\left[\rho \left( \bar{T} e^{i\int_{\tau_0}^{\tau} H^I_{\rm int}(\tau') d\tau'}\right) \hat{\cal{R}}^I_{\vec{k}_1}(\tau)\hat{\cal{R}}^I_{\vec{k}_2}(\tau)\hat{\cal{R}}^I_{\vec{k}_3}(\tau)   \left( Te^{-i\int_{\tau_0}^{\tau} H^I_{\rm int}(\tau') d\tau'}\right) \right]= \nonumber \\ &=& {\rm Tr}[\rho \hat{\cal{R}}^I_{\vec{k}_1}(\tau)\hat{\cal{R}}^I_{\vec{k}_2}(\tau)\hat{\cal{R}}^I_{\vec{k}_3}(\tau)] -i \int_{\tau_0}^{\tau} d\tau' \, {\rm Tr}\big[\rho \, [  \hat{\cal{R}}^I_{\vec{k}_1}(\tau)\hat{\cal{R}}^I_{\vec{k}_2}(\tau)\hat{\cal{R}}^I_{\vec{k}_3}(\tau) ,H^I_{\rm int}(\tau')] \big] + {\cal{O}}(H^2_{\rm int}) \nonumber \ ,\eea
where $T$ and $\bar{T}$ indicate the time and anti-time ordered product respectively,  the superscript $I$ indicates operators in the interaction picture, and $\tau_0$ is a conformal time about $60$ $e$-foldings before the end of inflation. Because comoving curvature perturbations $\hat{\cal{R}}^I_{\vec{k}}$ remain constant outside the horizon, we can evaluate the previous quantity when $\tau\to0$. Additionally, in the computation we assume that $\vec{k}_i\neq 0$. We will compute the two terms in the last line of the previous equation separately. 

The first term can be rewritten, by using the field redefinition (\ref{filedred}) and using the convolution theorem, as follows
\bea \label{aca} & & {\rm Tr}[\rho \hat{\cal{R}}^I_{\vec{k}_1}(0)\hat{\cal{R}}^I_{\vec{k}_2}(0)\hat{\cal{R}}^I_{\vec{k}_3}(0)] ={\rm Tr}[\rho \hat{\cal{R}}^I_{c,\vec{k}_1}(0)\hat{\cal{R}}^I_{c,\vec{k}_2}(0)\hat{\cal{R}}^I_{c,\vec{k}_3}(0)] + \\ \nonumber &+&\left(\frac{1}{2} \frac{\ddot{\phi}}{\dot{\phi} H}+\frac{1}{8 M_P^2}  \frac{\dot{\phi}^2}{H^2}\right) \left( \sum_{\vec{p}}{\rm Tr}[\rho \hat{\cal{R}}^I_{c,\vec{k}_1}(0)\hat{\cal{R}}^I_{c,\vec{k}_2}(0) \hat{\cal{R}}^I_{c,\vec{p}}(0)\hat{\cal{R}}^I_{c,\vec{k}_3-\vec{p}}(0)] + {\rm cyclic \ permut}\right) + \\  \nonumber &+&  \frac{1}{4 M_P^2}  \frac{\dot{\phi}^2}{H^2} \left(\frac{1}{k^2_3} \sum_{\vec{p}}  {\rm Tr}[\rho \hat{\cal{R}}^I_{c,\vec{k}_1}(0)\hat{\cal{R}}^I_{c,\vec{k}_2}(0) \hat{\cal{R}}^I_{c,\vec{p}}(0)\hat{\cal{R}}^I_{c,\vec{k}_3-\vec{p}}(0)] (\vec{k}_3-\vec{p})^2+ {\rm cyclic \ permut}\right) \ . \eea
The first term in the rhs of the previous equation vanishes, because of the form of $\rho$ [see the discussion after (\ref{rho6})]. The trace appearing in the second term can be written, using (\ref{modexp}), as
\bea \label{asa} & &\sum_{\vec{p}} {\rm Tr}[\rho \hat{\cal{R}}^I_{c,\vec{k}_1}(0)\hat{\cal{R}}^I_{c,\vec{k}_2}(0) \hat{\cal{R}}^I_{c,\vec{p}}(0)\hat{\cal{R}}^I_{c,\vec{k}_3-\vec{p}}(0)] = \sum_{\vec{p}} {\rm Tr}[\rho (A_{\vec{k}_1} {\cal R}_{k_1}(0)+A^{\dagger}_{-\vec{k}_1} {\cal R}^*_{k_1}(0) )\times \\ \nonumber &\times&(A_{\vec{k}_2} {\cal R}_{k_2}(0)+A^{\dagger}_{-\vec{k}_2} {\cal R}^*_{k_2}(0) )(A_{\vec{p}} {\cal R}_p(0)+A^{\dagger}_{-\vec{p}} {\cal R}^*_p(0) )(A_{\vec{k}_3-\vec{p}} {\cal R}_{|\vec{k}_3-\vec{p}|}(0)+A^{\dagger}_{-(\vec{k}_3-\vec{p})} {\cal R}^*_{|\vec{k}_3-\vec{p}|}(0) )] \ . \eea
There are six (non-vanishing) expectation values in the previous expression that we have to compute. We show two of the calculations, as examples:
\bea \label{zxz}  & & \sum_{\vec{p}}{\rm Tr}[\rho A_{\vec{k}_1} A_{\vec{k}_2}A^{\dagger}_{-\vec{p}}A^{\dagger}_{-(\vec{k}_3-\vec{p})}]  {\cal R}_{k_1}(0)  {\cal R}_{k_2}(0){\cal R}^*_{p}(0){\cal R}^*_{|\vec{k}_3-\vec{p}|}(0)=  \\ \nonumber &=&\left(2 \, {\rm Tr}[\rho A_{\vec{k}_1} A_{\vec{k}_2}A^{\dagger}_{\vec{k}_1}A^{\dagger}_{\vec{k}_2}]-\delta_{\vec{k}_1,\vec{k}_2}{\rm Tr}[\rho A_{\vec{k}_1} A_{\vec{k}_1}A^{\dagger}_{\vec{k}_1}A^{\dagger}_{\vec{k}_1}]\right)  |{\cal R}_{k_1}(0)|^2 |{\cal R}_{k_2}(0)|^2 \delta_{\vec{k}_1+\vec{k}_2+\vec{k}_3,0}\\ \nonumber &=& \left( 2 \,  {\rm Tr}[\rho (N_{\vec{k}_1}+1)(N_{\vec{k}_2}+1)]-\delta_{\vec{k}_1,\vec{k}_2} {\rm Tr}[\rho N_{\vec{k}_1} (N_{\vec{k}_1}+1)]\right)  |{\cal R}_{k_1}(0)|^2 |{\cal R}_{k_2}(0)|^2 \delta_{\vec{k}_1+\vec{k}_2+\vec{k}_3,0} \ ,\eea
and
\bea \label{zdz} \sum_{\vec{p}}{\rm Tr}[\rho A^{\dagger}_{-\vec{k}_1} A_{\vec{k}_2}A^{\dagger}_{-\vec{p}}A_{(\vec{k}_3-\vec{p})}]   {\cal R}^*_{k_1}(0)  {\cal R}_{k_2}(0){\cal R}^*_{p}(0){\cal R}_{|\vec{k}_3-\vec{p}|}(0)= \\ \nonumber = {\rm Tr}[\rho N_{\vec{k}_1}(N_{\vec{k}_2}+1)]|{\cal R}_{k_1}(0)|^2 |{\cal R}_{k_2}(0)|^2 \delta_{\vec{k}_1+\vec{k}_2+\vec{k}_3,0}\ . \eea
By computing the four remaining expectation values, one obtains the result for expression (\ref{asa}) as
\be \sum_{\vec{p}} \label{bab} {\rm Tr}[\rho \hat{\cal{R}}^I_{c,\vec{k}_1}(0)\hat{\cal{R}}^I_{c,\vec{k}_2}(0) \hat{\cal{R}}^I_{c,\vec{p}}(0)\hat{\cal{R}}^I_{c,\vec{k}_3-\vec{p}}(0)] = 2 \,F(\rho, \vec{k}_1 , \vec{k}_2) |{\cal R}_{k_1}(0)|^2 |{\cal R}_{k_2}(0)|^2 \delta_{\vec{k}_1+\vec{k}_2+\vec{k}_3,0} \ , \ee
where
\be F(\rho, \vec{k}_1 , \vec{k}_2)={\rm Tr}[\rho (2 N_{\vec{k}_1}+1) (2 N_{\vec{k}_2}+1)]-\delta_{\vec{k}_1,\vec{k}_2} {\rm Tr}[\rho N_{\vec{k}_1}(N_{\vec{k}_1}+1)]\ . \ee
In a similar way, we obtain the value of the last term on the rhs of expression (\ref{aca})
\bea \label{sas} & & \hspace{-2cm}   \frac{1}{k^2_3} \sum_{\vec{p}}  {\rm Tr}[\rho \hat{\cal{R}}^I_{c,\vec{k}_1}(0)\hat{\cal{R}}^I_{c,\vec{k}_2}(0) \hat{\cal{R}}^I_{c,\vec{p}}(0)\hat{\cal{R}}^I_{c,\vec{k}_3-\vec{p}}(0)] (\vec{k}_3-\vec{p})^2=\\ \nonumber &=&  \frac{k_1^2+k_2^2}{k_3^2}  F(\rho, \vec{k}_1 , \vec{k}_2) |{\cal R}_{k_1}(0)|^2 |{\cal R}_{k_2}(0)|^2 \delta_{\vec{k}_1+\vec{k}_2+\vec{k}_3,0} \ .\eea
Introducing (\ref{bab}) and (\ref{sas}) in (\ref{aca}), we obtain
\bea \label{bbb} & & {\rm Tr}[\rho \hat{\cal{R}}^I_{\vec{k}_1}(0)\hat{\cal{R}}^I_{\vec{k}_2}(0)\hat{\cal{R}}^I_{\vec{k}_3}(0)]=\\ & & \nonumber \frac{\delta_{({\tiny\sum} {\vec{k}_i}),0}  }{L^6} \left(\frac{H}{\dot{\phi}}\right)^4  \frac{H^4}{\prod_i (2 k_i)^3} \left[2  \frac{\ddot{\phi}}{\dot{\phi} H}  k_3^3+ \frac{\dot{\phi}^2}{H^2} \left( \frac{1}{2} k_3^3+\frac{1}{2} k_3 (k_1^2+k_2^2)\right)\right] F(\rho, \vec{k}_1 , \vec{k}_2)+{\rm cyclic\ permut} \eea
It remains to compute the second term in the last line of (\ref{pert}). It contains the expectation value of the commutator of three fields operators and the interaction Hamiltonian. Let us compute the first term of the commutator. By Fourier expanding the field operators appearing in the interaction Hamiltonian and using the convolution theorem twice, we have, at leading order in perturbations 
\bea \label{cvc}     -i \int_{\tau_0}^{0} d\tau'  \,  {\rm Tr}\big[\rho \hat{\cal{R}}^I_{\vec{k}_1}(0)\hat{\cal{R}}^I_{\vec{k}_2}(0)\hat{\cal{R}}^I_{\vec{k}_3}(0) H^I_{\rm int}(\tau') \big]=- i \int_{\tau_0}^{0} d\tau'a^3(\tau') \left(\frac{\dot\phi}{H}\right)^4 H M_P^{-2} L^3 \times  \\ \nonumber     \times \sum_{\vec{p}}  \sum_{\vec{k}'} \frac{1}{p^2} {\rm Tr}[\rho  \hat{\cal{R}}^I_{c,\vec{k}_1}(0)\hat{\cal{R}}^I_{c,\vec{k}_2}(0)  \hat{\cal{R}}^I_{c,\vec{k}_3}(0)\partial_{\tau'} \hat{\cal{R}}^I_{c,\vec{k}'}(\tau')\partial_{\tau'} \hat{\cal{R}}^I_{c,\vec{p}_-\vec{k}'}(\tau')\partial_{\tau'} \hat{\cal{R}}^I_{c,-\vec{p}}(\tau')] \ . \eea
The previous expressions contains 20 expectation values similar to (\ref{zxz}) and (\ref{zdz}), but now involving six creation and annihilation operators. By computing those expectation values and by taking into account that, at leading order in slow-roll parameters
\be {\cal{R}}_{k}(0)\partial_{\tau'}{\cal{R}}^*_{k}(\tau')=\frac{H^2}{2 k L^3 } \tau' \left(\frac{H}{\dot\phi}\right)^2 e^{i k \tau'} \ , \ee
(\ref{cvc}) reduces to 
\bea \label{nmn} &-i& \int_{\tau_0}^{0} d\tau' \, {\rm Tr}\big[\rho \hat{\cal{R}}^I_{\vec{k}_1}(0)\hat{\cal{R}}^I_{\vec{k}_2}(0)\hat{\cal{R}}^I_{\vec{k}_3}(0) H^I_{\rm int}(\tau') \big]=\delta_{({\tiny\sum} {\vec{k}_i}),0}  \frac{H^4}{4  L^6 } \left(\frac{H}{\dot\phi}\right)^2 \frac{M_P^{-2}}{k_1 k_2 k_3} \left(\frac{1}{k_1^2}+\frac{1}{k_2^2}+\frac{1}{k_3^2}\right) \nonumber  \\ & & \Bigg\{ \frac{1-e^{i (k_1+k_2+k_3)\tau_0 }}{k_1+k_2+k_3} \bigg[  {\rm Tr}[\rho (N_{\vec{k}_1}+1) (N_{\vec{k}_2}+1)(N_{\vec{k}_3}+1)]- \\   \nonumber & &  \hspace{4cm} -	\left(\frac{\delta_{\vec{k}_1,\vec{k}_2}}{2} {\rm Tr}[\rho N_{\vec{k}_1} (N_{\vec{k}_1}+1) (N_{\vec{k}_3}+1) ] + {\rm cyclic \ permut}\right) \bigg] +  \\   \nonumber & &
\bigg( \frac{1-e^{i (-k_1+k_2+k_3)\tau_0 }}{-k_1+k_2+k_3} \big[  {\rm Tr}[\rho N_{\vec{k}_1} (N_{\vec{k}_2}+1)(N_{\vec{k}_3}+1)] -\frac{\delta_{\vec{k}_2,\vec{k}_3}}{2} {\rm Tr}[\rho N_{\vec{k}_1} N_{\vec{k}_2} (N_{\vec{k}_2}+1) ] \big] \\   \nonumber & &  \hspace{11 cm}+ {\rm cyclic \ permut} \bigg) + \\   \nonumber & &
\bigg( \frac{1-e^{i (-k_1-k_2+k_3)\tau_0 }}{-k_1-k_2+k_3} \big[  {\rm Tr}[\rho N_{\vec{k}_1} N_{\vec{k}_2} (N_{\vec{k}_3}+1)]-	\frac{\delta_{\vec{k}_1,\vec{k}_2}}{2} {\rm Tr}[\rho N_{\vec{k}_1} (N_{\vec{k}_1}+1) (N_{\vec{k}_3}+1) ] \big] \\   \nonumber & &  \hspace{11cm}  + {\rm cyclic \ permut} \bigg)+ \\   \nonumber & &
\frac{1-e^{i (-k_1-k_2-k_3)\tau_0 }}{-k_1-k_2-k_3} \bigg[  {\rm Tr}[\rho N_{\vec{k}_1} N_{\vec{k}_2} N_{\vec{k}_3}]- 	\left(\frac{\delta_{\vec{k}_1,\vec{k}_2}}{2} {\rm Tr}[\rho N_{\vec{k}_1} (N_{\vec{k}_1}+1) N_{\vec{k}_3} ] + {\rm cyclic \ permut}\right) \bigg] \Bigg\} \ .\eea
The other term of the commutator can be computed in a similar way producing
\bea \label{gfg} &-i& \int_{\tau_0}^{0} d\tau' \,{\rm Tr} \big[\rho  H^I_{\rm int}(\tau') \hat{\cal{R}}^I_{\vec{k}_1}(0)\hat{\cal{R}}^I_{\vec{k}_2}(0)\hat{\cal{R}}^I_{\vec{k}_3}(0) \big]=  \delta_{({\tiny\sum} {\vec{k}_i}),0}  \frac{H^4}{4  L^6 } \left(\frac{H}{\dot\phi}\right)^2 \frac{M_P^{-2}}{k_1 k_2 k_3} \left(\frac{1}{k_1^2}+\frac{1}{k_2^2}+\frac{1}{k_3^2}\right) \nonumber \\ \nonumber & & \Bigg\{ \frac{1-e^{i (k_1+k_2+k_3)\tau_0 }}{k_1+k_2+k_3} \bigg[  {\rm Tr}[\rho N_{\vec{k}_1} N_{\vec{k}_2} N_{\vec{k}_3} ]- \\   \nonumber & &  \hspace{5cm} -	\left(\frac{\delta_{\vec{k}_1,\vec{k}_2}}{2} {\rm Tr}[\rho N_{\vec{k}_1} (N_{\vec{k}_1}+1) N_{\vec{k}_3}] + {\rm cyclic \ permut}\right) \bigg] +  \\   \nonumber & &
\bigg( \frac{1-e^{i (-k_1+k_2+k_3)\tau_0 }}{-k_1+k_2+k_3} \big[  {\rm Tr}[\rho (N_{\vec{k}_1}+1) N_{\vec{k}_2} N_{\vec{k}_3}] -\frac{\delta_{\vec{k}_2,\vec{k}_3}}{2} {\rm Tr}[\rho (N_{\vec{k}_1}+1) N_{\vec{k}_2} (N_{\vec{k}_2}+1) ] \big]+ \\    \nonumber & &  \hspace{11cm}+ {\rm cyclic \ permut}\bigg) + \\   \nonumber & &
\bigg( \frac{1-e^{i (-k_1-k_2+k_3)\tau_0 }}{-k_1-k_2+k_3} \big[  {\rm Tr}[\rho (N_{\vec{k}_1}+1) (N_{\vec{k}_2}+1) N_{\vec{k}_3}]-	\frac{\delta_{\vec{k}_1,\vec{k}_2}}{2} {\rm Tr}[\rho (N_{\vec{k}_1}+1) N_{\vec{k}_1} N_{\vec{k}_3} ]\big] + \\   \nonumber & &  \hspace{11cm} + {\rm cyclic \ permut}\bigg) + \\   \nonumber & &
\frac{1-e^{i (-k_1-k_2-k_3)\tau_0 }}{-k_1-k_2-k_3} \bigg[  {\rm Tr}[\rho (N_{\vec{k}_1}+1) (N_{\vec{k}_2}+1) (N_{\vec{k}_3}+1)]- \\    & &  \hspace{4cm} -	\left(\frac{\delta_{\vec{k}_1,\vec{k}_2}}{2} {\rm Tr}[\rho N_{\vec{k}_1} (N_{\vec{k}_1}+1) (N_{\vec{k}_3}+1) ] + {\rm cyclic \ permut}\right)\bigg] \Bigg\}\ .\eea
By subtracting (\ref{gfg}) from (\ref{nmn}), we have
\bea \label{opo}  & & \hspace{-2.9cm} -i \int_{\tau_0}^{0} d\tau' \, {\rm Tr}\big[\rho \, [  \hat{\cal{R}}^I_{\vec{k}_1}(0)\hat{\cal{R}}^I_{\vec{k}_2}(0)\hat{\cal{R}}^I_{\vec{k}_3}(0) ,H^I_{\rm int}(\tau')] \big]=\\ \nonumber &=&  \frac{\delta_{({\tiny\sum} {\vec{k}_i}),0} }{L^6} \left(\frac{H}{\dot{\phi}}\right)^4 \frac{H^4}{M_P^2} \frac{1}{\prod_i (2 k_i)^3} \left[ 4 \frac{\dot{\phi}^2}{H^2}  \sum_{i>j} k_i^2 k_j^2 
\right] G(\rho, \vec{k}_1,\vec{k}_2,\vec{k}_3) \ , \eea
where $G(\rho, \vec{k}_1,\vec{k}_2,\vec{k}_3)$ is a symmetric function under cyclic permutations of $(k_1,k_2,k_3)$ given by
\bea& &  \hspace{-1.5cm} G(\rho, \vec{k}_1,\vec{k}_2,\vec{k}_3)=\\ \nonumber &=&\frac{(1-\cos{( k_t \tau_0)})}{k_t} \bigg[ \bigg({\rm Tr}[\rho N_{\vec{k}_1} N_{\vec{k}_2}]+{\rm Tr}[\rho N_{\vec{k}_1}]+{\rm cyclic\ permut} \bigg)+1- \\ \nonumber &-&\left( \frac{\delta_{\vec{k}_1,\vec{k}_2}}{2}   {\rm Tr}[\rho N_{\vec{k}_1} (N_{\vec{k}_1} +1)] +{\rm cyclic\ permut} \right)  \bigg] +\\ \nonumber  &+& \bigg[ \frac{(1-\cos{( \tilde{k}_1 \tau_0)})}{\tilde{k}_1} \bigg({\rm Tr}[\rho N_{\vec{k}_1} N_{\vec{k}_2}]+{\rm Tr}[\rho N_{\vec{k}_1} N_{\vec{k}_3}]-{\rm Tr}[\rho N_{\vec{k}_2} N_{\vec{k}_3}]+{\rm Tr}[\rho N_{\vec{k}_1}] +\\ \nonumber &+& \frac{\delta_{\vec{k}_2,\vec{k}_3}}{2}   {\rm Tr}[\rho N_{\vec{k}_2} (N_{\vec{k}_2}+1)] \bigg) +{\rm cyclic\ permut}  \bigg]  \eea
with $k_t\equiv k_1+k_2+k_3$ and $\tilde{k}_i\equiv k_t-2 k_i$. 

Putting (\ref{bbb}) and (\ref{opo}) together and taking into account (\ref{delta}) and (\ref{newpowspect}), we obtain

\bea \label{vbv} & &{\rm Tr}[\rho \hat{{\cal{R}}}_{\vec{k}_1}(0)\hat{{\cal{R}}}_{\vec{k}_2}(0)\hat{{\cal{R}}}_{\vec{k}_3}(0)]=\delta_{({\tiny\sum} {\vec{k}_i}),0}  \  P^{\rho}_{\cal{R}}(\vec{k}_1) P^{\rho}_{\cal{R}}(\vec{k}_2)\times \\ \nonumber &  & \left[\frac{1}{2} \left(3\epsilon-2\eta+\epsilon \frac{k_1^2+k_2^2}{k_3^2}\right) f(\rho, \vec{k}_1 , \vec{k}_2)+4\epsilon \frac{k_1^2 k_2^2}{k_3^3} g(\rho, \vec{k}_1 , \vec{k}_2,\vec{k}_3) \right] + {\rm cyclic\  permut} \ ,  \eea
where we have defined
 \be  f(\rho, \vec{k}_1 , \vec{k}_2)\equiv  \frac{F(\rho, \vec{k}_1 , \vec{k}_2)}{{\rm Tr}[\rho (2 N_{\vec{k}_1}+1)]{\rm Tr}[\rho (2 N_{\vec{k}_2}+1) ]}\ ,\ee
\be g(\rho, \vec{k}_1 , \vec{k}_2,\vec{k}_3)\equiv \frac{G(\rho, \vec{k}_1 , \vec{k}_2,\vec{k}_3)}{{\rm Tr}[\rho (2 N_{\vec{k}_1}+1)]{\rm Tr}[\rho (2 N_{\vec{k}_2}+1) ]} \ . \ee
Expression (\ref{vbv}) is the three-point function that we presented in section \ref{bispec}.


\begin{thebibliography}{99}

\bibitem{parker-toms} Parker L. and Toms D.J., {\it Quantum field theory in curved spacetime: quantized fields
and gravity}, Cambridge University Press, (2009).

\bibitem{parkerthesis} Parker L., {\it The creation of particles in an expanding universe}, Ph.D. thesis, Harvard University (1966). 

\bibitem{parker68-69-71} Parker L., {\it Phys. Rev. Lett.} {\bf 21} 562 (1968); {\it Phys. Rev.} {\bf 183}, 1057 (1969);  {\it Phys. Rev. D} {\bf 3}, 346 (1971).

\bibitem{hawk1}
Hawking S.W. , {\it Comm. Math. Phys.} {\bf 43}, 199 (1975).

\bibitem{inflation} Guth A.,  {\it Phys. Rev.} D{\bf 23}, 347 (1981).
%
Starobinsky A.A., {\it Phys. Lett.} B{\bf 91}, 99 (1980).
%
Linde A.D., {\it Phys. Lett.} B{\bf 108}, 389 (1982); {\it Phys. Lett.} B{\bf 129}, 177 (1983).
%
Albrecht A. and  Steinhardt P. J., {\it Phys. Rev. Lett.} {\bf 48},1220 (1982).
%
Sato K.,  {\it Mon. Not. Roy. Astron. Soc.} {\bf 195}, 467 (1981).
%
\bibitem{inflation2} Mukhanov V.F. and Chibisov G.V., {\it JETP Letters} {\bf33}, 532(1981).
%
Hawking S. W., {\it Phys. Lett.} B{\bf 115}, 295 (1982).
%
Guth A. and Pi S.-Y., {\it Phys. Rev. Lett.} {\bf 49}, 1110 (1982).
%
Starobinsky A. A., {\it Phys. Lett.} B{\bf 117}, 175 (1982).
%
Bardeen J.M., Steinhardt P.J. and Turner M.S., {\it Phys. Rev.} D {\bf 28}, 679 (1983).

\bibitem{wmap7} Komatsu E. {\it et al.}, {\it Seven-Year Wilkinson Microwave Anisotropy Probe (WMAP) Observations: Cosmological interpretation} arXiv:1001.4538.

\bibitem{whitepaper} Komatsu E. {\it et al.}, {\it Non-gaussianity as a probe of the physics of the primordial universe and the astrophysics of the low redshift universe}, arXiv:0902.4759.

\bibitem{Weinberg2008} Weinberg S., \textit{Cosmology}, Oxford University Press, (2008).

\bibitem{dodelson} Dodelson S., \textit{Modern Cosmology}, Academic Press, (2003).

\bibitem{liddlelyth} Liddle A. R., and Lyth D.H., \textit{Cosmological inflation and large-scale structure}, Cambridge University Press, (2000).
 
\bibitem{baumann} Baumann D., {\it TASI Lectures on inflation}, arXiv:0907.5424.
 
\bibitem{bunch-davies} Bunch T.S. and Davies P.C.W., {\it Proc. Roy. Soc.} {\bf A}360, 117 (1978).

\bibitem{gasperiniveneziano} Gasperini M., Giovannini M., and Veneziano G., Phys. Rev. D {\bf 48}, R439 (1993).

\bibitem{bhattacharya}  Bhattacharya K., Mohanty  S., and Rangarajan R., {\it Phys. Rev. Lett.} {\bf 96} 121302 (2006).  Bhattacharya K., Mohanty S., and Nautiyal A., {\it Phys. Rev. Lett.} {\bf 97} 251301 (2006). Das S. and Mohanty S.,  {\it Phys. Rev. D} {\bf 80} 123537 (2009).

\bibitem{magueijo} Ferreira P. G. and Magueijo J.,  {\it Phys. Rev. D} {\bf 78} R061301 (2008).

\bibitem{boyanovsky} Boyanovsky D., de Vega H. J., and Sanchez N. G.,  {\it Phys. Rev. D} {\bf 78} 123006 (2006).

\bibitem{holman} Holman R. and Tolley A., {\it JCAP}05(2008)001.

\bibitem{meerburg} Meerburg P. D., van der Schaar J. P., and Corasanati P. S., JCAP05(2009)018; JCAP02(2010)001.

\bibitem{parkerfulling73} Parker L.  and Fulling S. A., {\it Phys. Rev. D} {\bf 7} 2357 (1973).

\bibitem{komatsuthesis} Komatsu E., {\it The pursuit of non-gaussianities fluctuations in the cosmic microwave background}, astro-ph/020639.

\bibitem{komatsuspergel}  Komatsu E. and Spergel D.N., {\it Phys. Rev. D} {\bf 63} 063002 (2001).

\bibitem{creminellizaldarriaga} Creminelli P. and  Zaldarriaga M., {\it JCAP}10(2004)006.

\bibitem{vonneumann}  von Neuman J., {\it Mathematical foundations of quantum mechanics}, Princeton University Press, Princeton, New Jersey (1955).


\bibitem{anderson} Anderson P. R.,  Eaker W., Habib S. H., Molina-Paris C., and Motola E., {\it Phys. Rev. D} {\bf 62} 12019 (2000); Anderson P. R., Molina-Paris C. , and Motola E., {\it Phys. Rev. D} {\bf 72} 043515 (2005).

\bibitem{messiah} Messiah A., {\it Quantum Mechanics}, North-Holland Publishing Company, Amsterdam, (1961), p. 334.


\bibitem{parkernature} Parker L., {\it Nature} {\bf 261} 20-23 (1976).

\bibitem{Glenz-Parker09}  Glenz M. and  Parker L., {\it Phys. Rev. D} {\bf 80} 063534 (2009).

\bibitem{Parker07} Parker L.,  {\em Amplitude of perturbations from inflation}, hep-th/0702216.

\bibitem{agulloetal09} Agullo I., Navarro-Salas J.,  Olmo G.J. and  Parker L., {\it Phys. Rev. Lett.} {\bf 103}, 061301 (2009); {\it Gen. Rel. Grav.} {\bf 41}, 2301 (2009). 

\bibitem{agulloetal10PRD} Agullo I., Navarro-Salas J.,  Olmo G.J. and  Parker L., {\it Phys. Rev. D.} {\bf 81}, 043514 (2010).

\bibitem{Mukhanov86} Mukhanov V.S., {\it JETP Lett.} {\bf 41}, 493 (1986). Sasaki S. {\it Prog. Theor. Phys.} {\bf 76}, 1036 (1986).

\bibitem{copietal} Copi J., Huterer D., Schwarz D. J., and Starkman G. D., {\it Large angle anomalies in the CMB}, arXiv:1004.5602. 

\bibitem{maldacena}  Maldacena J., {\it JHEP}05(2003)013.

\bibitem{ganckomatsu}  Ganc J. and Komatsu E., {\it A new method for calculating the primordial bispectrum in the squeezed limit}, arXiv:1006.5457. 

\end{thebibliography}
\end{document}